\documentclass[10pt]{article}  
\pdfoutput=1
\usepackage{setspace}
\usepackage{ifpdf}
\usepackage{wrapfig, blindtext}
\usepackage{hhline}
\usepackage{enumitem}
\usepackage{titlesec}
\titlelabel{\thetitle.\quad}
\usepackage{amsmath}    
\usepackage{graphicx}   
\usepackage{verbatim}   
\usepackage{color}     
\usepackage{subcaption}
\usepackage{xfrac}
\usepackage[symbol]{footmisc}
\usepackage{fullpage}
\usepackage[super,comma]{natbib}
\usepackage{hyperref}   
\usepackage[top=1in, bottom=1in, left=1.15in, right=1.15in]{geometry}
\usepackage{setspace}
\usepackage{pdflscape}
\usepackage{multirow}
\usepackage{verbatim}
\usepackage{changepage}
\usepackage{algpseudocode}
\usepackage{algorithm}
\usepackage{newcent}
\usepackage{float}
\floatstyle{plaintop}
\restylefloat{table}
\usepackage{cases}
\usepackage{hyperref}
\usepackage{wasysym}
\usepackage{braket}
\usepackage[utf8]{inputenc}
\usepackage[T1]{fontenc}
\usepackage{todo}

\title{Modeling the Expected Performance of the REgolith X-ray Imaging Spectrometer (REXIS)} 

\author{Niraj K. Inamdar$^{a}$\footnote{Corresponding author. Email: inamdar@mit.edu}, Richard P. Binzel$^{a}$, Jae Sub Hong$^{b}$, Branden Allen$^{b}$,\\ 
Jonathan Grindlay$^{b}$, Rebecca A. Masterson$^{c}$\\
$^{a}$\small{Massachusetts Institute of Technology, Department of Earth, Atmospheric}\\
\small{and Planetary Sciences, Cambridge, MA} \\
$^{b}$\small{Harvard-Smithsonian Center for Astrophysics, Cambridge, MA}\\
$^{c}$\small{Massachusetts Institute of Technology, Space Systems Laboratory, Cambridge, MA}}
\date{}
\begin{document} 
\maketitle 

\begin{abstract}
OSIRIS-REx is the third spacecraft in the NASA New Frontiers Program and is planned for launch in 2016. OSIRIS-REx will orbit the near-Earth asteroid (101955) Bennu, characterize it, and return a sample of the asteroid’s regolith back to Earth. The Regolith X-ray Imaging Spectrometer (REXIS) is an instrument on OSIRIS-REx designed and built by students at MIT and Harvard. The purpose of REXIS is to collect and image sun-induced fluorescent X-rays emitted by Bennu, thereby providing spectroscopic information related to the elemental makeup of the asteroid regolith and the distribution of features over its surface.

Telescopic reflectance spectra suggest a CI or CM chondrite analog meteorite class for Bennu, where this primitive nature strongly motivates its study. A number of factors, however, will influence the generation, measurement, and interpretation of the X-ray spectra measured by REXIS.  These include: the compositional nature and heterogeneity of Bennu, the time-variable Solar state, X-ray detector characteristics, and geometric  parameters for the observations.

In this paper, we will explore how these variables influence the precision to which REXIS can measure Bennu's surface composition. By modeling the aforementioned factors, we place bounds on the expected performance of REXIS and its ability to ultimately place Bennu in an analog meteorite class. 
\end{abstract}


\section{Introduction}\label{sec:intro} 
In 2016, NASA is scheduled to launch OSIRIS-REx (``\textbf{O}rigins \textbf{S}pectral \textbf{I}nterpretation \textbf{R}esource \textbf{I}dentification \textbf{S}ecurity \textbf{R}egolith \textbf{Ex}plorer''), a mission whose goal is to characterize and ultimately return a sample of the near-Earth asteroid (101955) Bennu (formerly 1999 RQ$_{36}$ and 
hereafter Bennu)\cite{ORExReview}. Bennu was chosen as the target asteroid for OSIRIS-REx for several reasons. Spectral similarities in different near-infrared bands to B-type asteroids 24 Themis and 2 Pallas raise the intriguing possibility that Bennu is a transitional object between the two. Furthermore, Bennu's reflectance spectra suggest that 
it may be related to a CI or CM carbonaceous chondrite analog meteorite class \cite{clark2011asteroid}. Carbonaceous chondrites are believed to be amongst the most primitive material in the Solar System, undifferentiated and with refractory elemental abundances very similar to the Sun's. The discovery of water on the surface of 24 Themis provides additional scientific motivation for studying Bennu. 

Bennu belongs to a class of asteroids known as near-Earth asteroids (NEA). Its semimajor axis is roughly $1~\textrm{AU}$, and its orbit crosses Earth's \cite{campins2010origin}. While this makes Bennu a particularly accessible target for exploration, it also makes Bennu a non-negligible impact risk to Earth. Calculations of Bennu's orbital elements suggest an impact probability of $\sim 10^{-4}-10^{-3}$ by the year 2182, which coupled with its relatively large size (mean radius $\sim 250~\textrm{m}$) makes Bennu one of the most hazardous asteroids known \cite{milani2009long}. 

\begin{figure}[htpb]
\begin{center}
\includegraphics[width=0.8\textwidth]{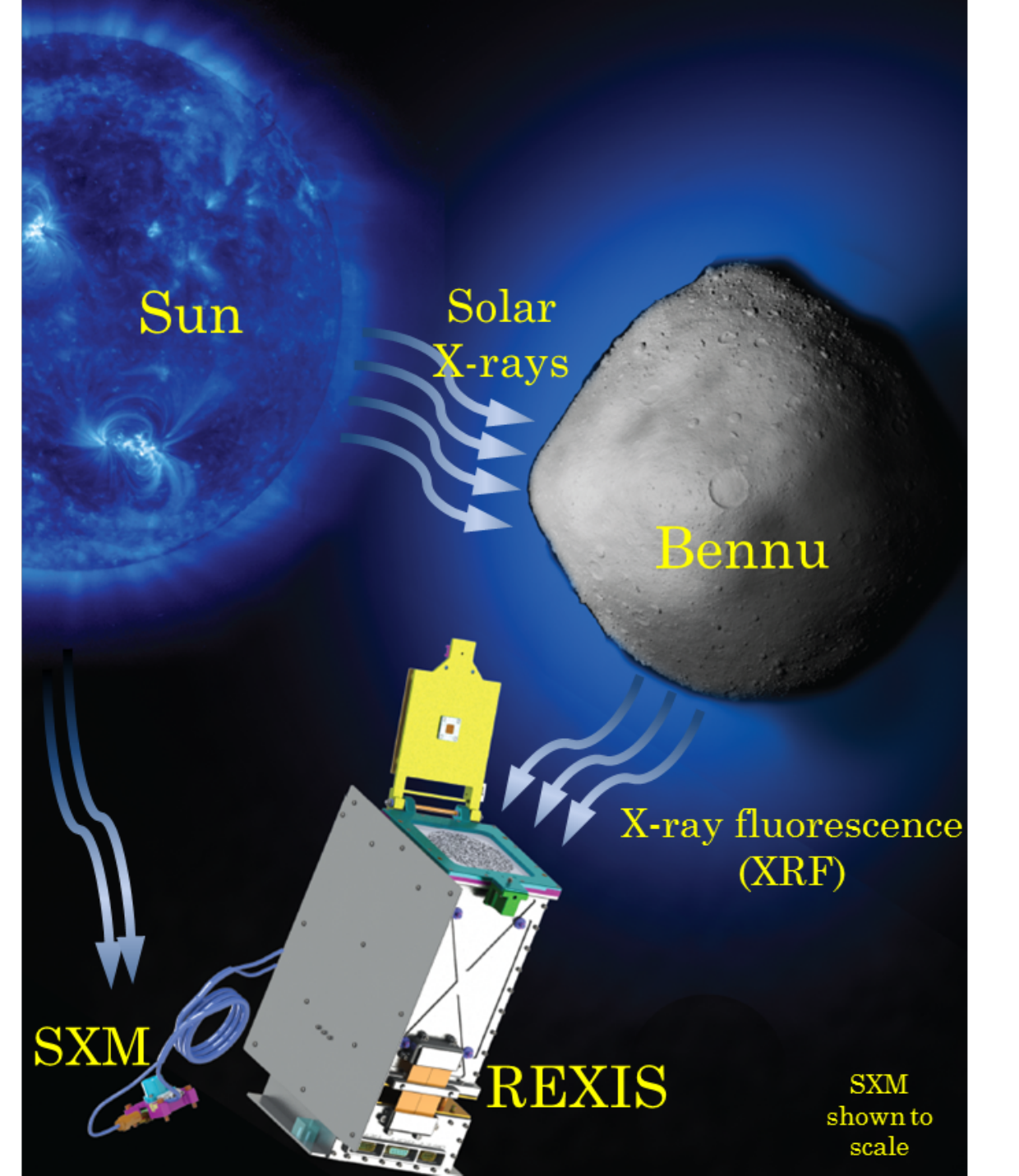}
\caption[]{REXIS principle of operation demonstrated schematically. Except for the Solar X-ray Monitor, which is shown to 
scale relative to REXIS, the rest of the figure is not to scale. X-rays from the Sun impinge the regolith of Bennu, giving
rise to X-ray fluorescence. These X-rays enter REXIS, where they are collected by CCDs. The radiation cover (shown in
yellow) serves as a shade to prevent Solar radiation from entering REXIS. At the same time, the Solar X-ray Monitor 
(SXM), which is mounted on a different surface of OSIRIS-REx, collects Solar X-rays directly in order to understand the 
time variance of the Solar X-ray spectrum. The OSIRIS-REx spacecraft is not shown.}
\label{fig:REXIS_PoO} 
\end{center}
\end{figure}

Taken together, these unique features make Bennu an attractive target for future study. In order to better characterize Bennu's composition and physical state, OSIRIS-REx is equipped with a suite of instruments, amongst which is REXIS. REXIS, a student experiment aboard OSIRIS-REx, is an X-ray imaging spectrometer (``\textbf{RE}golith \textbf{X}-ray \textbf{I}maging \textbf{S}pectrometer'') whose purpose is to reconstruct elemental abundance ratios of Bennu's regolith by measuring X-rays fluoresced by Bennu in response to Solar X-rays (Fig. \ref{fig:REXIS_PoO})\cite{allen2013regolith,JonesSmithREXIS}. More details regarding REXIS's systems-level organization and operation can be found in Jones, \textit{et al.} \cite{JonesSmithREXIS}.

\subsection{Description of REXIS}

\begin{figure}[htpb]
\begin{center}
\fbox{\includegraphics[width=0.95\textwidth]{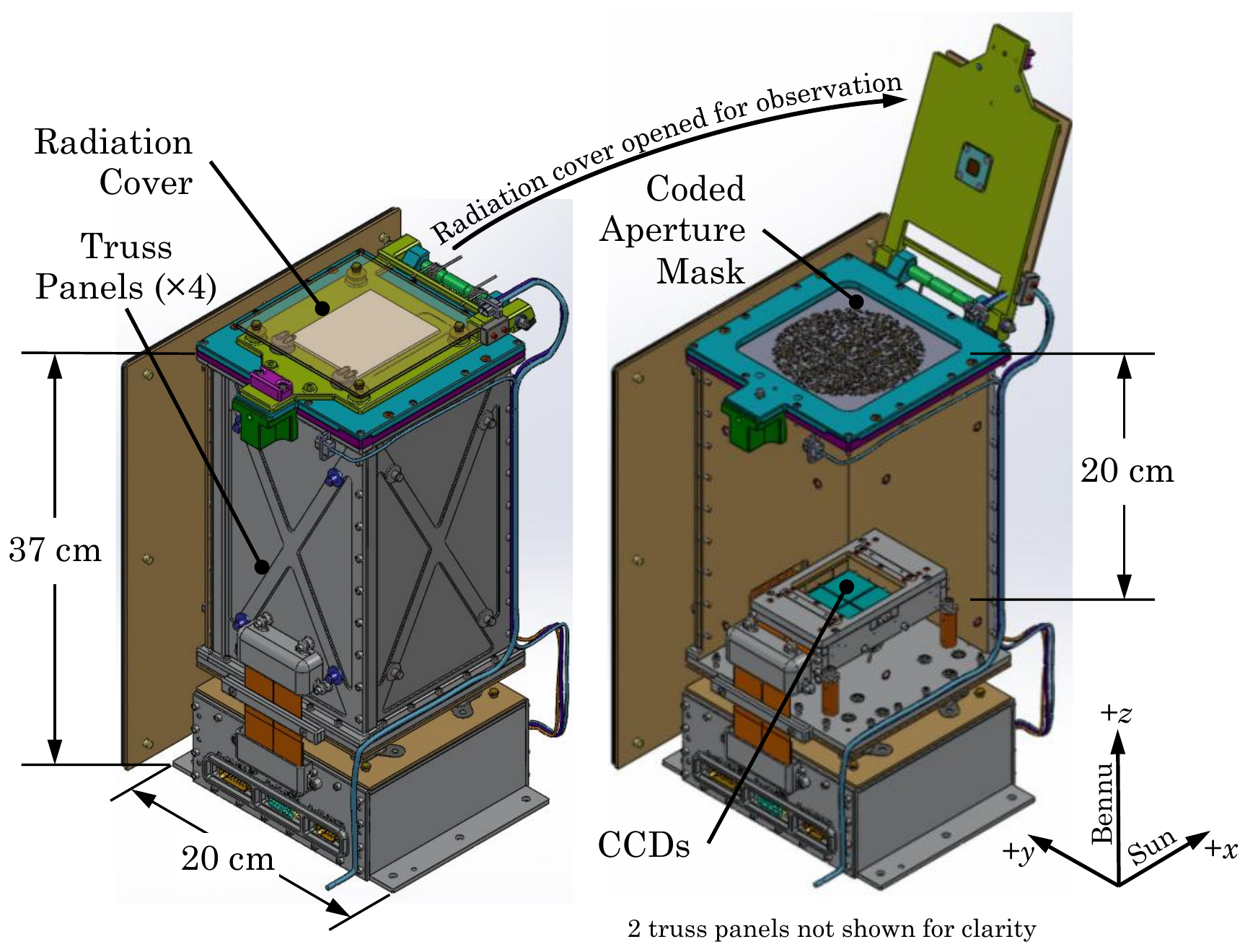}}
\caption[REXIS geometry.]{REXIS geometry. On the left, we show the instrument while it is in its stowed position,
with the radiation cover closed in order to protect the CCDs from radiation damage. On the right, we show the 
instrument in its observing state, with the radiation cover opened to allow the CCDs to collect X-rays. The coded
aperture mask is mounded above the truss structure and allows for fine angular resolution detection of local elemental
abundances in Imaging Mode. In Collimator Mode, however, the telescopic ``focal length'' of 20 cm and the diameter of
the coded aperture mask (9.84 cm) provide the coarse angular resolution. In Spectral Mode, which we focus on in this
paper, the surface area of the CCDs, the diameter of the coded aperture mask, and the focal length drive the instrument
grasp (see Sec. \ref{sec:grasp}).}
\label{fig:REXIS_geom}
\end{center}
\end{figure}

REXIS is comprised by two distinct, complementary instruments. The first is the primary spectrometer. Measuring approximately 37 cm high and 20 cm wide, it is mounted on the main instrument deck of OSIRIS-REx and houses 4 charged coupled devices (CCDs) that measure X-rays emitted by the Bennu's regolith (Fig. \ref{fig:REXIS_geom}). REXIS images X-rays by means of a coded aperture mask mounted atop the spectrometer tower. The X-ray shadow pattern cast by the mask on the detector plane and knowledge of the mask pattern allows for a reprojection of the measured X-rays back onto the asteroid, so that localized enhancements in the X-ray signal on roughly 50 m scales can be identified on Bennu's surface. During the mission cruise phase, a radiation cover protects the CCDs from bombardment by nonionizing radiation (such as Solar protons) that can create charge traps in the CCDs and subsequently degrade the detector resolution\cite{RadDamChandra}. This radiation cover is opened prior to calibration and asteroid observations (see below).

REXIS will observe Bennu for an overall observation period of $\sim 400$ hours. During this time, OSIRIS-REx will be in a roughly circular orbit along the asteroid's terminator with respect  to the Sun and about 1 km from the asteroid barycenter. REXIS will also collect calibration data. Since OSIRIS-REx orbits Bennu at approximately 1 km from the asteroid barycenter, and thus has a field of view that extends beyond the asteroid limb, cosmic sources of X-rays are a potential source of noise. Therefore, prior to asteroid observation, REXIS will observe the cosmic X-ray background (CXB) for a total of 3 hours. Furthermore, a period of 112 hours will be devoted to internal calibration to determine sources of X-ray noise intrinsic to the instrument itself. Throughout the operational lifetime of REXIS, a set of internal $^{55}$Fe radiation sources (which decay via electron capture to $^{55}$Mn with a primary intensity centered at $5.89~\mathrm{keV}$) will be used to calibrate the CCD gain.

The asteroid X-ray spectrum measured by REXIS depends on both the elemental abundances of the asteroid regolith and the Solar state at the time of measurement. In order to remove this degeneracy, a secondary instrument is required to measure Solar activity. The Solar X-ray Monitor (SXM), which is mounted on the Sun-facing side of REXIS, measures Solar activity and hence performs this function. The SXM contains a silicon drift diode (SDD) detecting element manufactured by Amptek, and generates a histogram of the Solar X-ray spectrum over each 32 s observational cadence. The Solar X-rays collected by the SXM allow for a time-varying reconstruction of the Solar state, so that, in principle, the only unknowns during interpretation of the asteroid spectrum are the regolith elemental abundances. The elemental abundances that we infer from the collected spectra are then used to map Bennu back to an analog meteorite class. 

\begin{figure}[htpb]
\centering
\fbox{\includegraphics[width=0.8\textwidth]{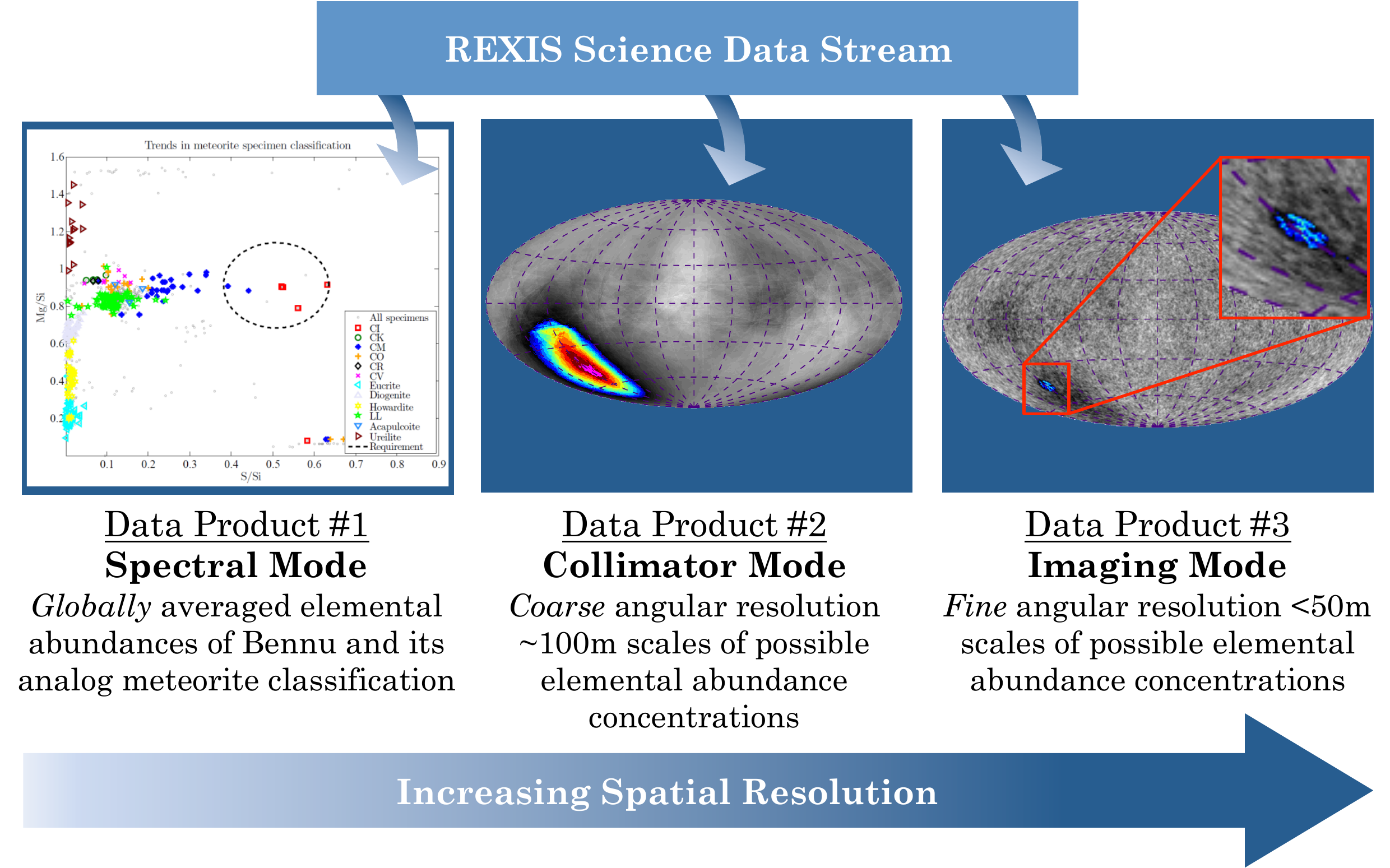}}
\caption[]{REXIS data processing modes. All data processing modes rely on the same data set, indicated schematically
above. In this paper, we focus on modeling the baseline performance of Spectral Mode, in which the 
globally-averaged spectrum of Bennu is used without regard to spatial variations of spectral features on the surface in
order to place Bennu within an analog meteorite class.}
\label{fig:REXIS_DataPipe} 
\end{figure}

During the REXIS observation period, X-rays emitted by Bennu are collected on board by CCDs (CCID-41s manufactured by MIT Lincoln Laboratory). The spectra that are generated from these data are then used to interpret the elemental abundance makeup of the asteroid. The baseline CCD data flow in a single stream, and REXIS data are processed in three distinct ``modes'':
(Fig. \ref{fig:REXIS_DataPipe}). These are:
\begin{description}[nolistsep,noitemsep]
	\item[Spectral Mode.] Only the overall accumulation of spectral CCD data over the instrument's observational period	are considered. No attempt is made at producing local elemental abundance or abundance ratio maps. Instead, the data are used to determine the average composition of the asteroid from the spectral data collected in order to correlate Bennu to a meteorite class of similar composition. 
	\item[Collimator Mode.] Coarse spatially resolved measurements of elemental abundances on the surface of Bennu are carried out in collimator mode using time resolved spectral measurements combined with the instrument attitude history and field of view (FOV) response function. The FOV response function is uniquely determined by the instrument focal length as well as the diameter and open fraction of the coded aperture mask.
	\item[Imaging Mode.] Higher spatial resolution spectral features on the asteroid surface are identified by applying	coded aperture imaging\cite{Caroli}. In each time step, the data are the same as in collimator mode, though the distribution of counts on the detector plane is reprojected (using the known mask pattern and an appropriate  deconvolution technique) onto the asteroid surface.
\end{description}
Science processing modes occur on the ground. Here, we are concerned with the performance of REXIS in Spectral Mode; discussion of REXIS's performance in imaging and collimator mode may be found in Allen, \textit{et al.}\cite{allen2013regolith}

\subsection{Placing Bennu Within an Analog Meteorite Class}
\begin{figure}[htpb]
        \centering
        \begin{subfigure}[t]{0.68\textwidth} 
                \includegraphics[width=\textwidth]{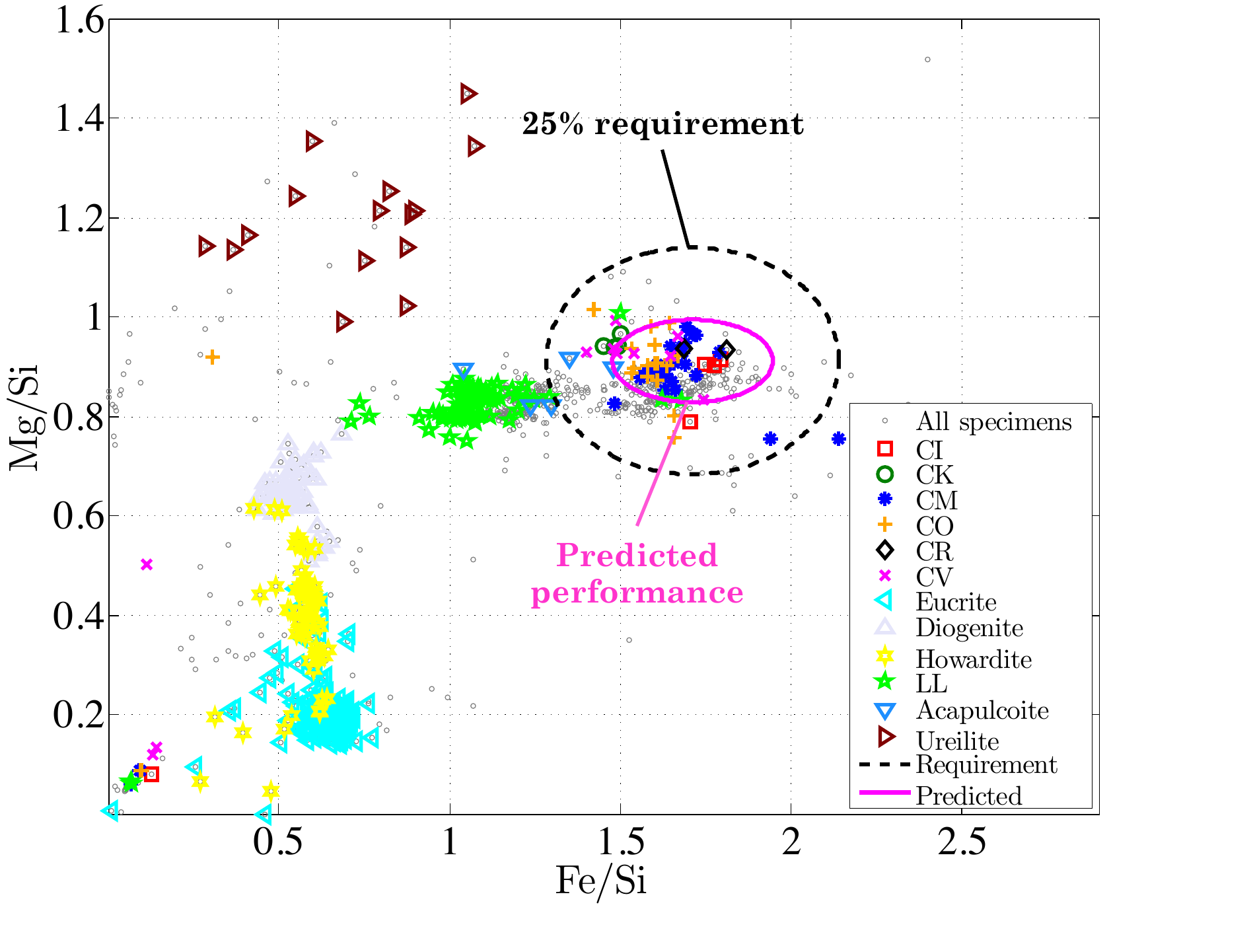}
                \label{FeSi_MgSi}
        \end{subfigure}%
        \quad
        \begin{subfigure}[t]{0.68\textwidth} 
                \includegraphics[width=\textwidth]{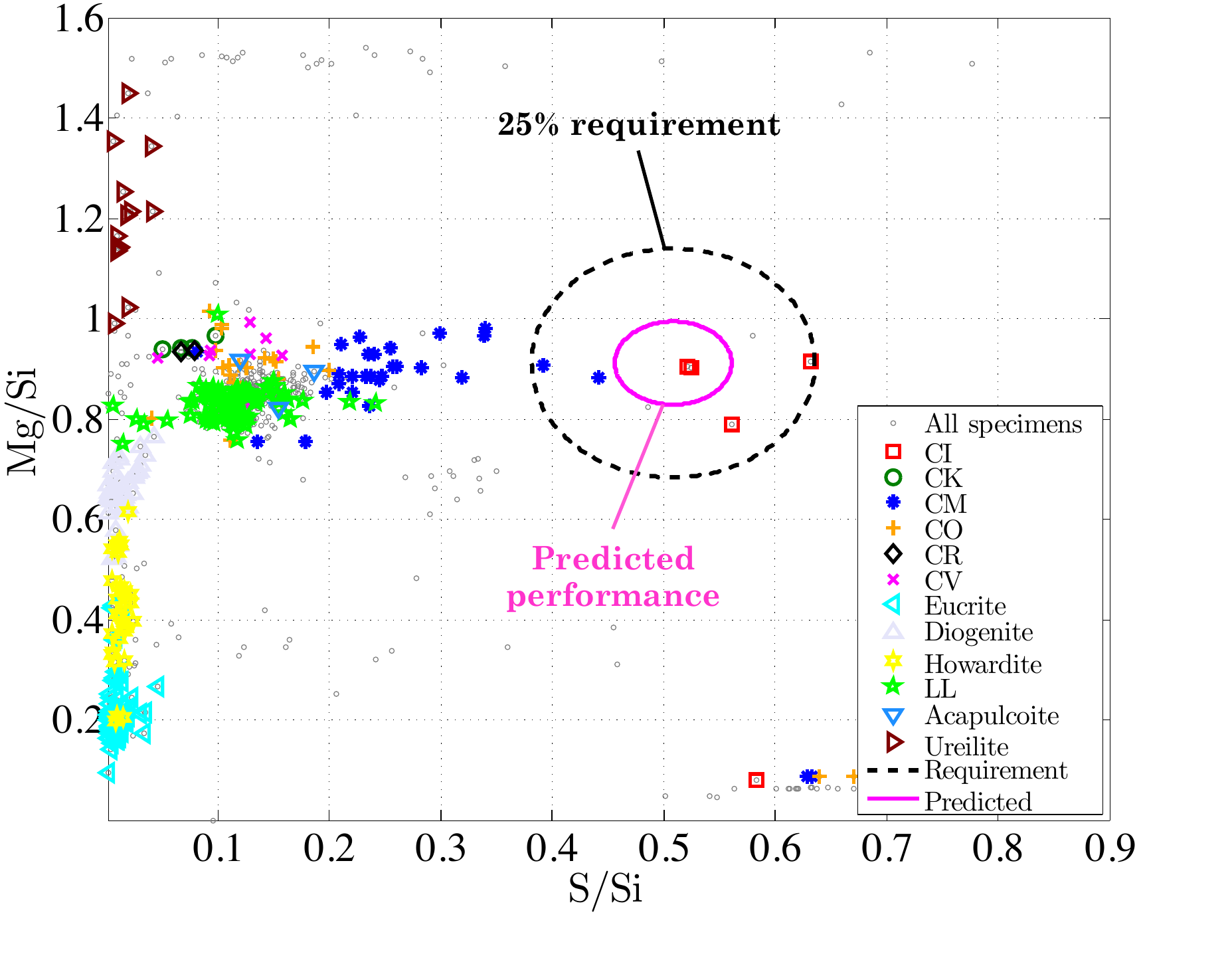}
                \label{SSi_MgSi}
        \end{subfigure}
        \caption{Trends in meteorite classification as a function of elemental abundance ratios. In the upper panel, we see 
        how  achondritic and chondritic meteorite specimens can be distinguished on the basis of their Mg/Si and 
        Fe/Si elemental abundance ratios\cite{NittlerData}. To differentiate between various chondrite subtypes, we rely 
        on the Mg/Si and S/Si ratios (lower panel). In both panels, we show the REXIS requirement of 25\% error centered 
        around a CI chondrite-like baseline. The expected REXIS performance under nominal conditions is indicated in the
        magenta ellipses. These error ellipses represent systematic error; further consideration of statistical error places the 
        confidence in these calculations of systematic error at $3.5\sigma$ (see Sec. \ref{sec:Results}). Composition data
        for meteorites are from Nittler, \textit{et al.} \cite{NittlerData}.
        }
        \label{fig:Nittler}
\end{figure}

One of the goals of REXIS is to place Bennu within an analog meteorite class. Meteorites of similar class can often be grouped based on chemical or isotopic similarity. In particular, it has been recognized that major chondritic and achondritic meteorite groups can be distinguished on the basis of elemental abundance ratios, as can various subchondritic types \cite{NittlerData}. In Fig. \ref{fig:Nittler}, we show how various meteorite classes can be grouped on the basis of elemental abundance ratios of Fe/Si, Mg/Si, and S/Si. REXIS therefore collects X-rays between energies of 0.5 and 7.5 keV, within which prominent Fe, Mg, S, and Si emission features are found. The particular X-ray energies associated with these elements are summarized in Table \ref{tab:SummaryOfElLines}. Consistent with the measurement of the X-ray signatures of these elements, REXIS has two high-level requirements associated with its performance in Spectral Mode. These are
\begin{itemize}[nolistsep,noitemsep]
\item REX-3: REXIS shall be able to measure the global ratios of Mg/Si, Fe/Si, and S/Si of Bennu within 25\% for that of a CI chondrite illuminated by a 4 MK, A3.3 Sun.
\item REX-6: REXIS shall meet performance requirements given no less than 420 hours of observation time of Bennu.
\end{itemize} 	
The first reflects the fact that REXIS must measure the stated elemental abundance ratios to within 25\% those of a typical CI chondrite during the quiet Sun.  25\% error is sufficient to distinguish between achondritic and chondritic types, as well as amongst various chondrite types, as indicated in Fig. \ref{fig:Nittler} by the dashed line ellipses. The second requirement reflects the fact that REXIS must attain its science objectives within its allotted observation period. 
\begin{table}[htpb]
	\centering
	\caption{Summary of lines of interest and their energies\cite{xdb2001}. In some cases, due to the close proximity of 
	spectral features to one another, they are combined with one another in the analysis below.}
		\begin{tabular}{|c|c|c|}\hline
Line Designation 			& Energy center [eV] 	& Notes 												\\ \hline\hline
Fe-L$\alpha$				& 705.0				& Due to proximity, is combined with Fe-L$\beta$ 				\\ \hline
Fe-L$\beta$ 				& 718.5 				& Due to proximity, is combined with Fe-L$\alpha$ 				\\ \hline
Mg-K$\alpha_1$/K$\alpha_2$	& 1,253.60 			& Due to proximity, is combined with Mg-K$\beta$ 				\\ \hline
Mg-K$\beta$	        			& 1,302.2 				& Due to proximity, is combined with Mg-K$\alpha_1$/K$\alpha_2$  	\\ \hline
Si-K$\alpha_1$/K$\alpha_2$	& 1,739.98/1,739.38 		& --- 													\\ \hline
 S-K$\alpha_1$/K$\alpha_2$	& 2,307.84/2,306.64 		& --- 													\\ \hline
		\end{tabular}
	\label{tab:SummaryOfElLines}
\end{table}

Our purpose in this work is to determine, for a given asteroid regolith composition, Solar state, and instrument characteristics, the spectrum that we expect to collect from the asteroid and the impact of data collection and processing on the eventual reconstruction of the hypothetical elemental abundances of Bennu. We model the expected performance of REXIS in its Spectral Mode and place bounds on its ability to place Bennu within an analog meteorite class. We will accomplish this in several steps. First, we model the ideal X-ray spectra that we expect to be generated by Bennu and the Sun. We then model the instrument response for both the spectrometer and the SXM, accounting for factors such as total throughput, detector active area, quantum efficiency, and spectral broadening. We then model the data processing. Here, we combine the instrument response-convolved spectra from both Bennu and the Sun to determine how well we can reconstruct Bennu's elemental abundances and place the asteroid within an analog meteorite class. We show that REXIS can accomplish its required objectives with sufficient margin.

\section{Methodology}\label{sec:Method}

\begin{figure}[htpb]
\begin{center}
\fbox{\includegraphics[width=.95\textwidth]{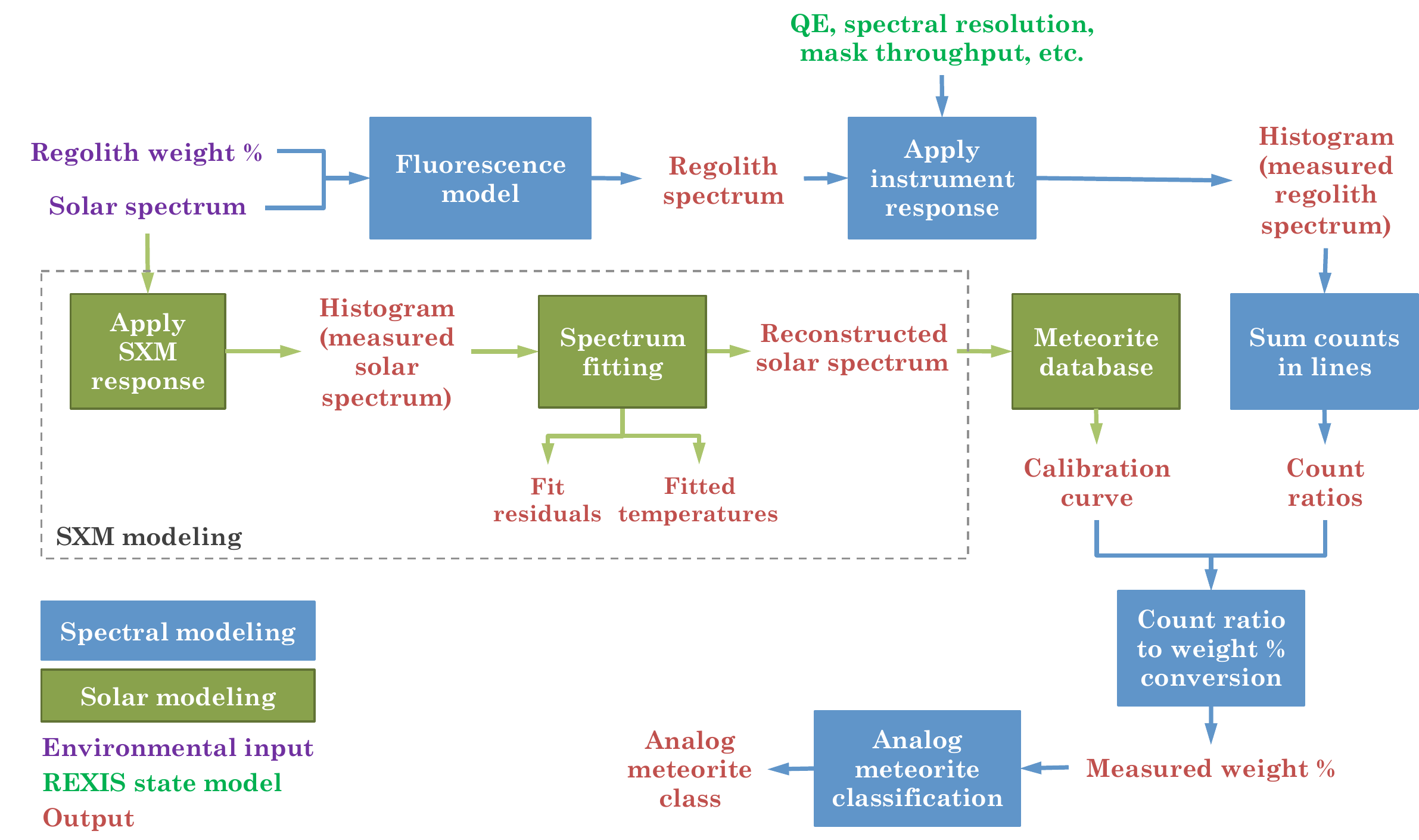}}
\caption[Spectral mode processing pipeline.]{Spectral mode simulation pipeline. In this flow diagram, we indicate the 
various processes contained in our simulations. The primary inputs may be broadly characterized  as environmental 
inputs (regolith composition, Solar spectrum; purple text) and REXIS state inputs, which includes geometric factors and 
detector characteristics accounting for the overall instrument response function (green text). Environmental inputs are used 
to create a synthetic regolith spectrum, to which the instrument response is then applied. Data processing is then 
simulated. In the SXM simulation, the ``actual'', differential emission measure-integrated quiet Solar spectrum is convolved 
with the SXM response function. We use the resulting histogram, which simulates the actual SXM data product, to 
reconstruct a best fit isothermal Solar spectrum. The best fit Solar spectrum, which is not detector response-convolved,
is then used to generate calibration curves using a meteorite database. These calibration curves map observed photon
count ratios to asteroid regolith weight ratios, which we then use to identify Bennu amongst analog meteorite classes.} 
\label{fig:Spect_Pipeline}
\end{center}
\end{figure}

Our overall methodology in simulating the expected performance of REXIS is summarized in Fig. \ref{fig:Spect_Pipeline}. Our basic procedure is to first simulate physical observables---in our case, asteroid and Solar spectra---under expected conditions. We then simulate the process of data collection for both the spectrometer and the SXM. Finally, we simulate the interpretation of the data and assess our ability to reconstruct the original observables using our processed data.  In order to assess our expected performance, throughout the entire modeling process, we keep track of all simulated quantities, including those that would be unknowns during the mission lifetime, such as the actual Solar and asteroid spectrum. 

\subsection{Simulating Observables} 
The baseline observables for the spectrometer and the SXM are the asteroid and Solar X-ray spectra, respectively. For the discussion that follows, we denote the asteroid spectrum $I_{\mathrm{B}}(E)$ and the Solar spectrum $I_{\astrosun}(E)$. The cosmic X-ray background spectrum, which we must also consider, is denoted $I_{\mathrm{CXB}}(E)$. In each case, the spectrum is a function of energy $E$ and has units of $\mathrm{photons/cm^2/s/Sr/keV}$. Based on ground observations, the expected asteroid spectrum $I_{\mathrm{B}}$ is that from a CI-like asteroid regolith. Since the OSIRIS-REx mission occurs during the Solar minimum, the expected Solar spectrum is that from a quiet Sun. 

\subsubsection{Asteroid Spectrum} 
Asteroid spectra are calculated using the standard fluorescence equation for the intensity of the fluorescent lines\cite{JenkinsQXS}. We also include contributions from coherent scattering.\footnote{All X-ray data, including fluorescent line energies, fluorescence yields, jump ratios, relative intensities, photoabsorption cross sections, and scattering cross sections are derived from the compilations of Elam, \textit{et al.}\cite{ELAM} and Kissel \cite{Kissel}. The Kissel scattering cross section data may be found at the following URLs:
\begin{itemize}[nolistsep,noitemsep]
\item[--] \url{http://ftp.esrf.eu/pub/scisoft/xop2.3/DabaxFiles/f0_rf_Kissel.dat}
\item[--] \url{http://ftp.esrf.eu/pub/scisoft/xop2.3/DabaxFiles/f0_mf_Kissel.dat}
\item[--] \url{http://ftp.esrf.eu/pub/scisoft/xop2.3/DabaxFiles/f1f2_asf_Kissel.dat}
\item[--] \url{http://ftp.esrf.eu/pub/scisoft/xop2.3/DabaxFiles/f0_EPDL97.dat}
\end{itemize}
}
The contribution from incoherent scattering is at least an order of magnitude less than that from coherent scattering and is ignored here \cite{LimNitt1}. We assume the asteroid, which is modeled as a sphere of radius 280 m, is viewed in a circular terminator orbit 1 km from the asteroid center. From the point of view of REXIS, half of the asteroid is illuminated while the other half is dark. Furthermore, the asteroid is not uniformly bright on its Sun-facing side, and the energy-integrated flux peaks at a point offset from the asteroid nadir. The effect of these angles is taken into account when generating the asteroid spectrum (for more details, see Appendix \ref{sec:AsteroidSec}). The asteroid spectrum itself is a function of the Solar spectrum. It is also, to a much lesser extent, a function of  the CXB, which is significantly lower in intensity than the incident Solar radiation, and which is only effective at inducing fluorescence at energies much higher than we are concerned with. In generating $I_{\mathrm{B}}(E)$, we use $I_{\astrosun}(E)$, as discussed below in Sec. \ref{sec:SolarSpec}.

\subsubsection{Solar Spectrum}\label{sec:SolarSpec} 
We calculate Solar X-ray spectra $I_{\astrosun}(E)$ using the CHIANTI atomic database\cite{CHIANTI_1,CHIANTI_2} and SolarSoftWare package \cite{SSW_pack}. The Solar spectrum is that generated by the Solar corona, the primary source of X-rays from the Sun. Since REXIS will be observing Bennu during the Solar minimum, we model the expected Solar spectrum by using the quiet Sun differential emission measure (DEM) derived from the quiet Sun data of Dupree, \textit{et al.}\cite{Dupree} and elemental abundances of Meyer \cite{Meyer,Anders}. 

The DEM is a quantity that encodes the plasma temperature dependence of the contribution function and hence intensity of the radiation \cite{LimNitt1,SolarCorona,LandiDEM}. The DEM can be derived from observations, and for the quiet Sun it tends to peak at a single temperature (in the range of about $3-6~\mathrm{MK}$), so that to first order, the Solar corona can be approximated as comprising an isothermal plasma. In general, however, the actual Solar X-ray spectrum will require an integration of the DEM over all temperatures present in the plasma along the observer's line of sight (see Appendix \ref{sec:SolarModelSec}). For higher coronal temperatures, access to higher energy states leads to a so-called hardening of the Solar spectrum \cite{LimNitt1}, an effect which is most pronounced during a Solar flare. In this case, the DEM peaks at more than one temperature. Since we expect the majority of our observations to take place while the Sun is relatively inactive, during data processing, we take advantage of the fact that the corona can be approximated as 
isothermal (for more details, see Appendix \ref{sec:SolarModelSec}). Finally, we note that the Solar X-ray spectrum depends on the elemental abundances of the Solar corona, for which several models are available \cite{LimNitt1}. However, our results are relatively insensitive to the coronal elemental abundance model employed.

\subsubsection{CXB Spectrum}\label{sec:CXBSpec} 
The CXB spectrum that we use in our models is calculated following Lumb, \textit{et al}\cite{LumbCXB}. In this model, $I_{\mathrm{CXB}}(E)$ is calculated by assuming that the CXB comprises two optically thin components\cite{MEKAL} and a power law component \cite{Zombeck}. In general, the CXB flux becomes comparable to the asteroid flux at $\sim 2~\mathrm{keV}$, near the S-K complex (see Fig. \ref{fig:Spect_Comp}). 
Measurement of sulfur is critical, since it enables us to differentiate amongst different chondritic varieties (Fig. \ref{fig:Nittler}). Hence, we ultimately find that measurement of the S/Si ratio is most sensitive to this particular source of noise and requires the longest amount of measurement time to achieve statistical significance (see Sec. \ref{sec:ObsTime}). 

\subsubsection{Internal Background}\label{sec:IntBack} 
\begin{figure}[htpb]
        \centering
        \begin{subfigure}[b]{0.625\textwidth}
                \includegraphics[width=\textwidth]{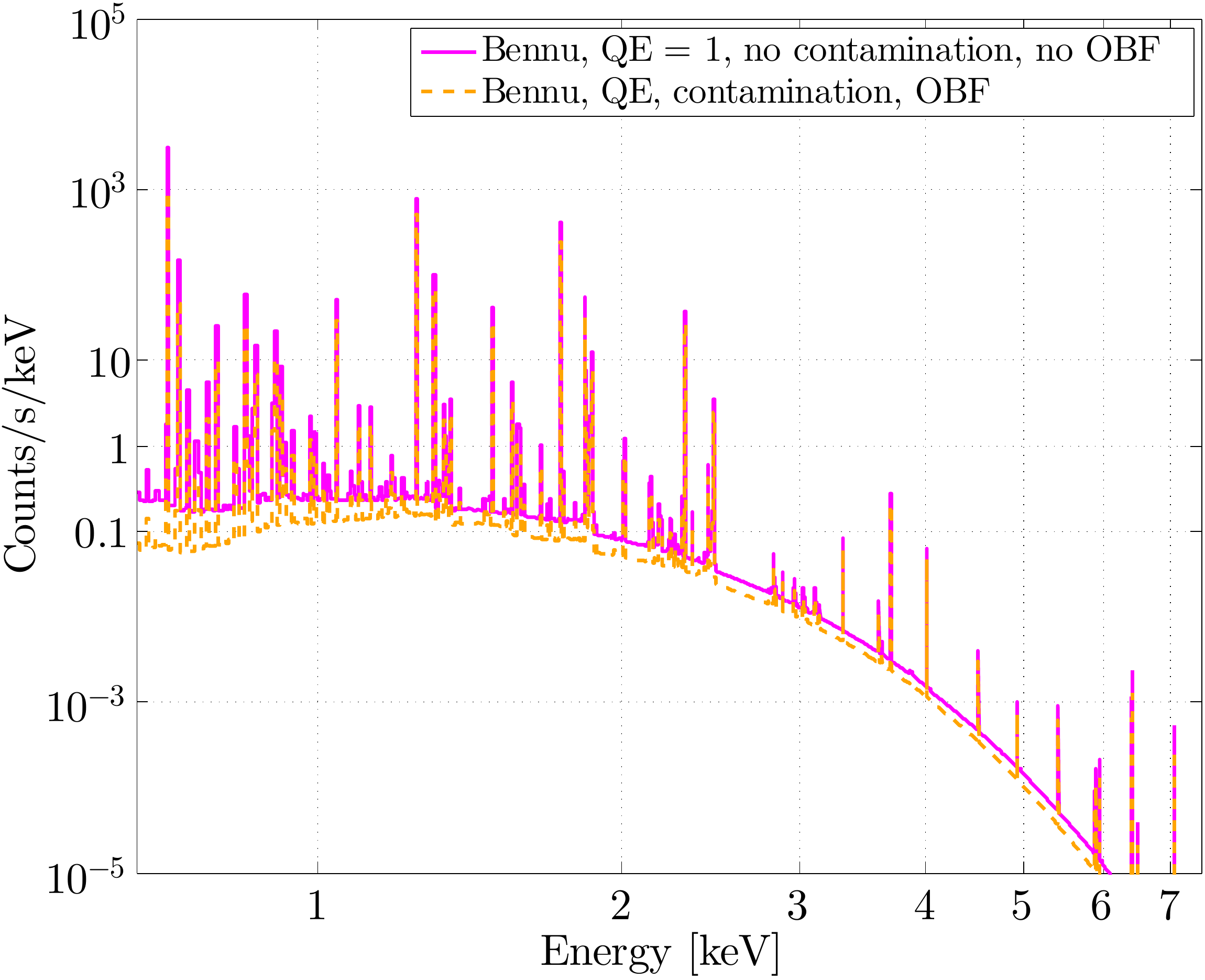}
                \label{fig:FeSi_MgSi}
        \end{subfigure}%
        \\~
        \begin{subfigure}[b]{0.625\textwidth}
                \includegraphics[width=\textwidth]{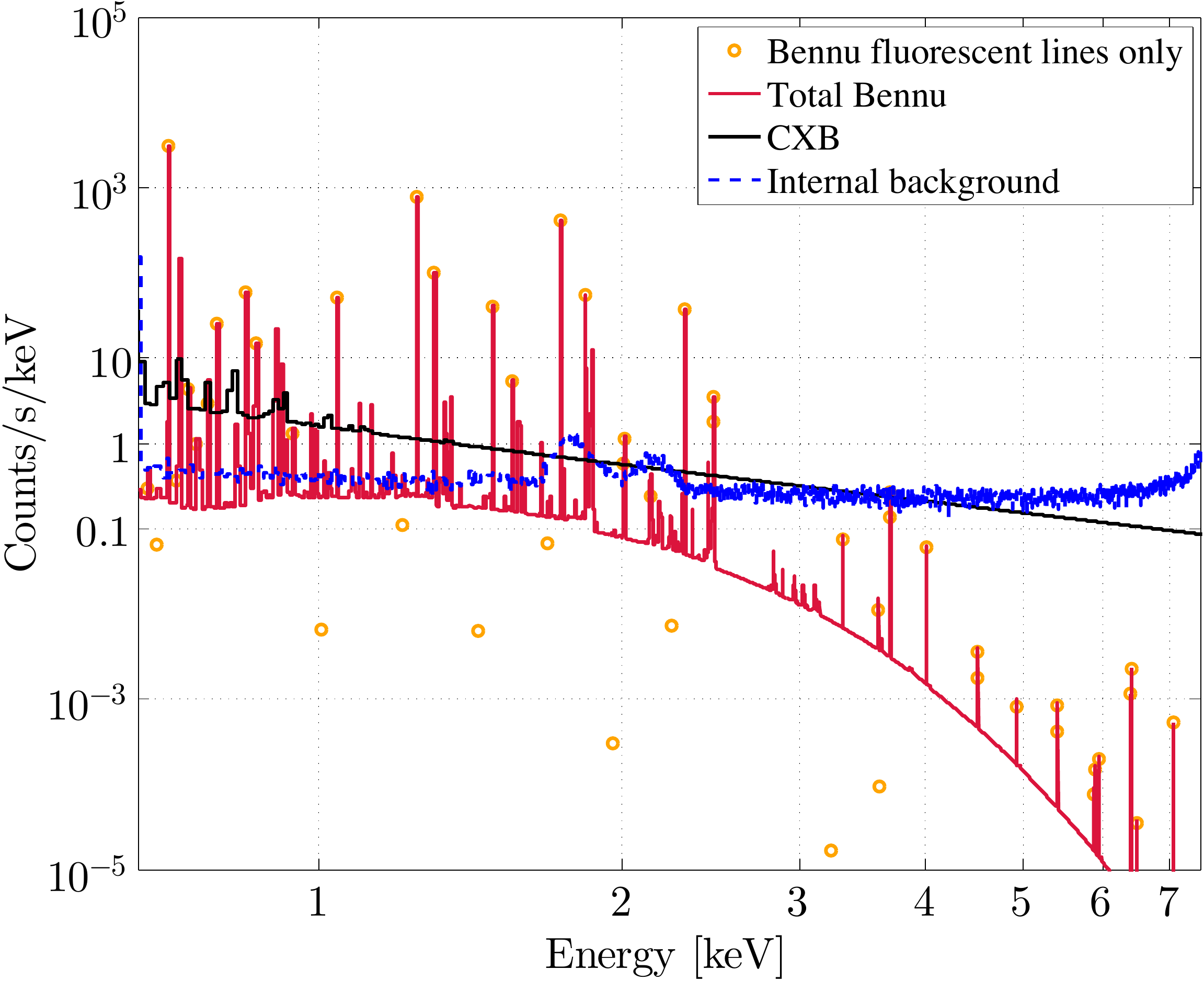}
                \label{fig:SSi_MgSi}
        \end{subfigure}
        \caption{
        Comparison of spectra of interest at an early stage of our model development. On the left panel, 
        we show our preliminary model of Bennu's spectrum as emitted directly from 
		the asteroid (solid magenta line). We also show Bennu's spectrum 
        after detector quantum efficiency, molecular contamination, and the optical blocking filter 
        (OBF; Fig. \ref{fig:InstResInputs}) are taken into account. On the right panel, we show 
        Bennu's spectrum compared to sources of noise. Fluorescent lines from Bennu 
        are shown as orange markers, while the complete Bennu spectrum, including scattering, 
        is shown in solid red. The cosmic X-ray background (CXB) is 
        shown in black, while internal noise (due to fluorescence from the instrument itself) is 
        shown with the dotted blue line. The Si-K complex from Bennu, which is the prominent set of 
        lines between 2 and 3 keV, is most strongly subject to the effects of noise. Internal background 
        has been scaled from Chandra data \cite{ACIS}. This model does not include oxygen as it falls 
        just below our model cut-off.
        }
        \label{fig:Spect_Comp}
\end{figure}

Fluorescence from the REXIS instrument itself can be present in the signal we measure. Incident X-rays primarily from Bennu (but also from the CXB) can strike the inner portions of the instrument and induce  fluorescence. Ideally, a ray-tracing simulation would be carried out to determine the extent of this internal noise. For our work, however, we use data from the Chandra ACIS instrument that has been suitably scaled down to match the detector area of the REXIS CCDs\cite{ACIS}. A comparison of Bennu's spectrum with that of the CXB and the internal background is shown in Fig. \ref{fig:Spect_Comp}.

\subsection{Instrument Response}
The next step after simulating the observables is to estimate how these will convolve with the instrument response. Thus we simulate the data collection process by applying the instrument response for both the  spectrometer and the SXM to our model spectrum. Inputs into the instrument response models include (along with the symbols that we use to denote each):

\begin{itemize}[nolistsep,noitemsep]
\item Observation time, $T_{\mathrm{obs}}$
\item Coded aperture mask throughput, $F$ (spectrometer only)
\item Grasp, $G = A_E \Omega$
	\begin{itemize}[nolistsep,noitemsep]
		\item Effective detector area, $A_E$
		\item Solid angle subtended by source with respect to detector, $\Omega$
	\end{itemize}	
\item Detector quantum efficiency, $Q(E)$ (a function of energy $E$)
\item Detector histogram bin width, $\Delta E$
\item Gain drift
\item Detector spectral resolution, $\mathrm{FWHM}$
\end{itemize}

In all cases where we are evaluating our results, we assume our measurements are well described within the realm of Poisson statistics.  The origin of the values used for each of these inputs varies. In sections below, we detail how each of these inputs is derived for our simulations. After the asteroid and Solar spectra have been convolved with the detector response functions, the basic output for each will be a histogram of photon counts as a function of energy. In Table \ref{tab:SpectObsInputs}, we summarize some of the major observational inputs into our simulations, while others are given in the text that follows.

\begin{table}[htp]
	\centering
	\caption[Observational inputs for instrument response]{Observational inputs for spectrometer 
	instrument response.
	}
\begin{tabular}{|c||c|} \hline
Parameter											& Value                 	\\ \hline \hline
Open fraction 										& 40.5\% 					\\ \hline
Histogram binning $\Delta E$ [$\mathrm{eV/bin}$]	& $\sim 15$ 	         	\\ \hline
Gain drift [$\mathrm{eV}$]							& $\pm 15$                	\\ \hline
Total observation time $T_{\mathrm{obs}}$ 			& 423 hours					\\ \hline
CXB calibration period $T_{\mathrm{CXB}}$ 			& 3 hours					\\ \hline
Internal background 								& \multirow{2}{*}{112 hours}\\
calibration period   $T_{\mathrm{int}}$				& 							\\ \hline
Solar state 										& Quiet Sun					\\ \hline
Regolith composition 								& $\sim$CI chondritic 		\\ \hline
\end{tabular}
	\label{tab:SpectObsInputs}
\end{table}

\subsubsection{Observation Time} 
The observation time, $T_{\mathrm{obs}}$, for the spectrometer is taken to be 423 hours. For the SXM, Solar spectra are recorded as histograms in 32 s intervals, roughly the time scale over which the Solar state can vary substantially. Time $T_{\mathrm{CXB}}$ and $T_{\mathrm{int}}$ is also allocated for CXB and internal calibration, respectively (see Sec. \ref{sec:BackSub}).

\subsubsection{Coded Aperture Mask Throughput}
The overall throughput of the instrument depends on the open fraction of the coded aperture mask (the fraction of open mask pixels to total mask pixels). For REXIS, nominally half of the coded aperture pixels are open. However, the presence of a structural grid network to support the closed pixels reduces the throughput further. Since the grid width is 10\% of the nominal pixel spacing, the count rate of photons incident on the REXIS detectors will be reduced by a factor $F = 0.5\times \left(1 - 0.1\right)^2 = 0.405$ due to the presence of the coded aperture mask.

\subsubsection{Grasp}\label{sec:grasp}
The grasp $G$, which has units of $\mathrm{cm^2Sr}$, is the quantity that encodes the solid angle $\Omega$ subtended by the target with respect to the detector, and the averaged detector geometric area $A_E$ 
that sees the target; detector efficiency is not accounted for in this term. The detector area does not comprise a single point, and since the field of view is not a simple cone, $G$ must in general be calculated numerically. We calculate $G$ for the CCDs using custom ray tracing routines in MATLAB and IDL. Since portions of the detector area can see the CXB that extends beyond the limb of the asteroid during observation, we keep track of this as well during our calculations\footnote{For a given differential area element on the detector surface, we calculate the solid angle subtended by Bennu and the CXB and multiply each by the differential area; we then average these values over the area of the detector.}. In Table \ref{tab:SpectGraspInputs}, we summarize $G$ for the CCDs, including individual contributions from Bennu and the CXB.

For the SXM, we assume that the solid angle subtended by the Sun is given by that for a distant source, $\pi\left(R_{\astrosun}/1~\mathrm{AU}\right)^2$ where $R_{\astrosun}$ is the Sun's radius. Since the Sun is located at such a distance that its incident rays can be treated as parallel, and since there are no structural elements driving the SXM viewing geometry substantially, we calculate $G$ for the SXM by simply multiplying the solid angle subtended by the Sun by the detector's $0.25~\mathrm{cm^2}$ area (Table \ref{tab:SXMInputs}).

\begin{table}[htp]
	\centering
	\caption[Geometric inputs for instrument response]{Geometric inputs for spectrometer instrument response during 
	primary observation period. These inputs assume a 280 m spherical asteroid radius,  average 1 km asteroid 
	centroid-to-spacecraft orbit, 9.84 cm mask coded area diameter, and a 20 cm focal length. During the calibration 
	period, when REXIS observes only the sky, the entire $4.24~\mathrm{cm^2 Sr}$ grasp is devoted to the CXB.
	}
\begin{tabular}{|c||c|c|c|} \hline
\multirow{2}{*}{Parameter}				& 					\multicolumn{3}{c|}{Value}			\\ \cline{2-4}
										& Bennu                 	& CXB 						& Total REXIS 	\\ \hline \hline
Averaged geometric 	 					& \multirow{3}{*}{15.16}	& \multirow{3}{*}{2.09} 	& \multirow{3}{*}{---}	\\ 
detector area							&							&			&				\\  
 $A_E$ [$\mathrm{cm^2}$]				&							&			&				\\ \cline{1-4}
Solid angle $\Omega$ [$\mathrm{Sr}$]	&	0.254					&	0.185	&	---			\\ \cline{1-4}
Grasp $G$ [$\mathrm{cm^2 Sr}$]			&	3.85					&	0.388	&	4.24		\\ \hline 
\end{tabular}
	\label{tab:SpectGraspInputs}
\end{table}
\begin{table}[htp]
	\centering
	\caption[Inputs for SXM response]{Inputs for SXM response.
	}
\begin{tabular}{|c||c|} \hline
Parameter											& Value                 		\\ \hline \hline
Histogram binning $\Delta E$ [$\mathrm{eV/bin}$]	& $\sim 30$ 	                \\ \hline
Single integration time 							& 32 s							\\ \hline
SDD area  $A_E$ [$\mathrm{cm^2}$]					& 0.25 							\\ \hline
Solid angle $\Omega$ [$\mathrm{Sr}$]     			& $\pi\left(R_{\astrosun}/1~\mathrm{AU}\right)^2 = 6.79 \times 10^{-5}$	\\ \hline
Grasp $G$ [$\mathrm{cm^2 Sr}$]						& $1.70\times 10^{-6}$			\\ \hline
\end{tabular}
	\label{tab:SXMInputs}
\end{table}
\subsubsection{Detector Quantum Efficiency}
The detector quantum efficiency, $Q(E)$, gives the overall reduction in counts registered by the detector due to absorption of incoming X-rays both by material overlying the CCDs and by the CCD material itself. In the case of the CCDs, we use the known material stackup\cite{BautzCCD} and widely-available photoabsorption cross section data\cite{hubbell1996tables} to determine the energy-dependent attenuation and hence quantum efficiency of the detector. We also include other possible sources of detection inefficiency, including built-up molecular contamination\cite{ACIS} and the optical blocking filter (OBF), which is a thin aluminum film deposited on the CCDs in order to prevent saturation from optical light. The combined contribution of all these to $Q(E)$ is shown in Fig. \ref{fig:InstResInputs}.\footnote{At the time of submission of this paper, some of the quantum efficiency data shown in Fig. \ref{fig:InstResInputs} is no longer up to date. However, the impact on our results are negligible, and future work will incorporate more accurate data. For more on the characterization of the REXIS CCDs, see Ryu, \textit{et al.}\cite{ryu2014development}} For the SXM, SDD efficiency curves are taken from manufacturer's data\cite{AMPTEK}.

\subsubsection{Detector Histogram Binning}
The histograms that are generated by the spectrometer data are binned in intervals of width $\Delta E$. Photons detected by the REXIS CCDs are assigned a 9 bit energy value, so that over an energy range of $0.5-7.5~\mathrm{keV}$, $\Delta E = 7~\mathrm{keV}/2^9 \sim 15~\mathrm{eV}$ (Table \ref{tab:SpectObsInputs}). For the SXM, there are 256 energy bins, so that $\Delta E \sim 30~\mathrm{eV}$ (Table \ref{tab:SXMInputs}).

\subsubsection{Gain Drift}\label{sec:GainDrift}
Our ability to accurately define line features depends on our ability to accurately calibrate the gain of the detectors. In the case of the spectrometer, we employ on-board $^{55}$Fe calibration sources in order to 
determine the line centers. The strength of the $^{55}$Fe sources has been chosen to ensure that within a given time period, the sources' line centers can be determined with $3\sigma$ accuracy to within one bin width. In our work, we shift the gain at $5.9~\mathrm{keV}$ randomly by $\pm~15~\mathrm{eV}$ for each simulation we perform. 

In the case of the SXM, we will use known Solar spectral features to accurately calibrate the gain over each integration period. Since the count rate for the SXM is so high, we can accurately determine line centers without counting statistics having too great an effect. 

\subsubsection{Detector Spectral Resolution}
\begin{figure}[htpb]
        \centering
        \begin{subfigure}[b]{0.625\textwidth}
                \includegraphics[width=\textwidth]{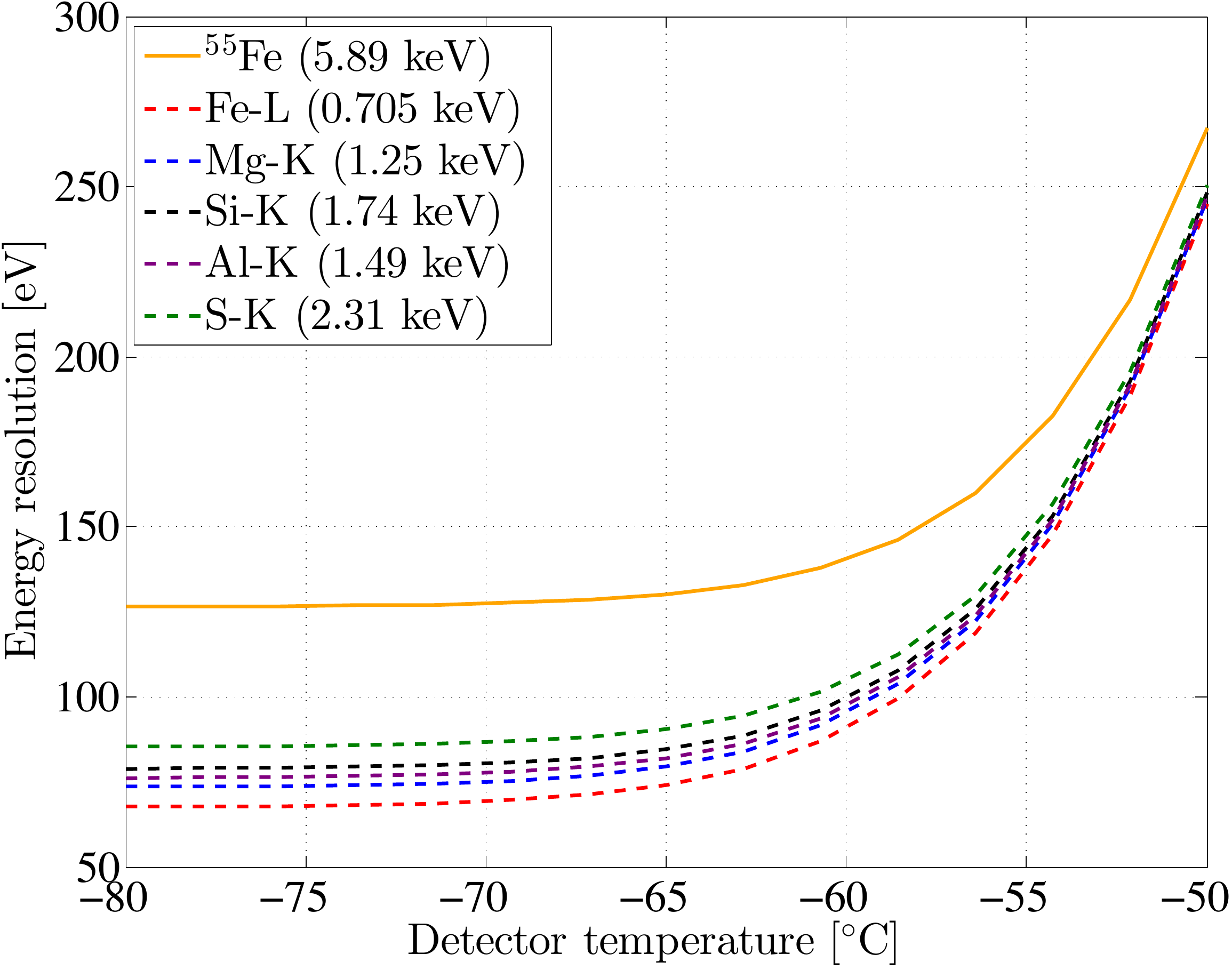}
                \label{fig:FWHM}
        \end{subfigure}%
	\\~
        \begin{subfigure}[b]{0.625\textwidth}
                \includegraphics[width=\textwidth]{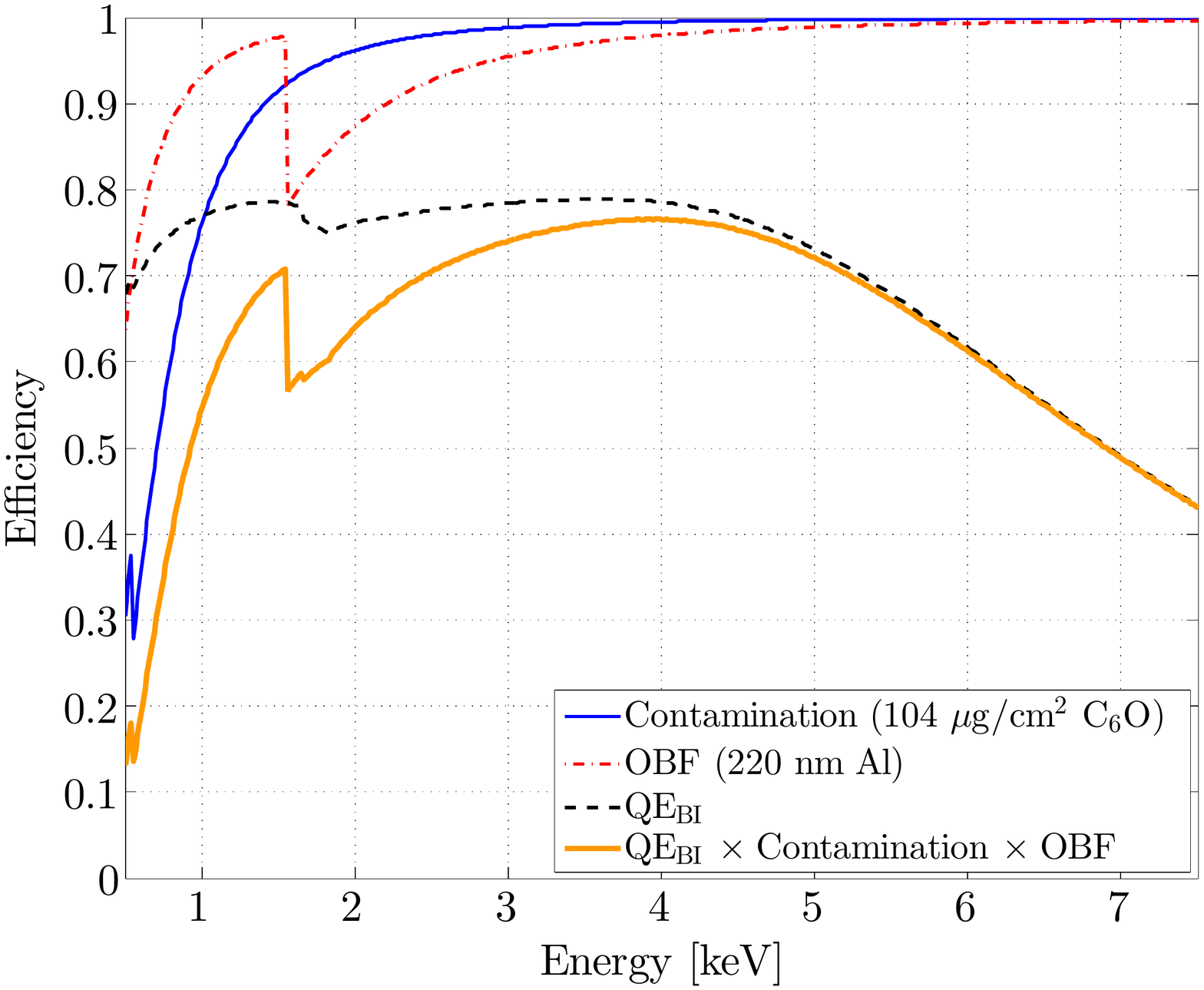}
                \label{fig:QE}
        \end{subfigure}
        \caption{Drivers of spectral resolution and instrument response. In the left panel, we show the energy resolution of 
        		the CCDs as a function of detector temperature for various energies. The energies indicated are those 
		associated with the line centers of spectral lines of interest. Initial tests have indicated that
                detector performance is at or near Fano-limited at the required detector operating temperature of 
                $-60~\mathrm{^\circ C}$. Curves have been calculated using the method given in Appendix 
                \ref{sec:FWHMCalc}.
                In the right panel, we show the quantum efficiency of the CCDs as a function of energy. Several sources of 
                efficiency degradation are indicated. Molecular contamination is indicated by the solid blue line. The effect of the 
                optical blocking filter, whose purpose is to attenuate optical light from Bennu that could cause saturation of the 
                detectors, is indicated by the dash-dotted red line. The dashed black line indicates the radiation attenuation 
                due to the composition stackup of the back-illuminated CCID-41. The total effect of all these are indicated by 
                the thick orange line. Note the quantum efficiency estimate for our back illuminated CCD includes a 
                conservative margin.
                }
        \label{fig:InstResInputs}
\end{figure}

The detector energy resolution, which we denote by $\mathrm{FWHM}$ (full width half maximum), describes the width of the Gaussian distribution that a delta function-like spectral line would assume due to broadening. Natural broadening, which is typically on the order of a few eVs, is negligible in comparison to broadening from the detector itself. For the CCDs, the $\mathrm{FWHM}$  is a function of both photon energy and detector temperature\cite{Suzaku}. The detector temperature drives dark current, which subsequently increases the width of the Gaussian. REXIS's required detector operating temperature $T$ is $-60~\mathrm{^\circ C}$ 
or below, while the predicted temperature at the time of writing is $\sim 20~\mathrm{^\circ C}$ colder than the requirement. Since the CCD temperature is the strongest driver of spectral resolution for a given line, and since the CCDs are passively cooled, in our results below, we calculate the performance over the range of detector temperatures between the requirement and prediction. CCD $\mathrm{FWHM}$ is determined using a combination of experimental data and analytical expressions, in a procedure outlined in Appendix \ref{sec:FWHMCalc}. Initial test results show that CCD performance is near or at Fano-limited. $\mathrm{FWHM}$ as a function of detector temperature for energies at the line centers of interest is shown in Fig. \ref{fig:InstResInputs}. For the SXM, the situation is somewhat simpler, since the SXM is cooled actively via a thermoelectric cooler. In this case, based on the manufacturer's test data, we have assumed that $\mathrm{FWHM}\left(5.9~\mathrm{keV}\right) = 125~\mathrm{eV}$.

\subsubsection{Calculating the Instrument Response Function}
In this section, we summarize how all the above inputs combine to generate the instrument response and a spectrum histogram. We denote the baseline intensity as $I_0(E)$ [= $I_{\mathrm{B}}(E)$, $I_{\astrosun}(E)$, or $I_{\mathrm{CXB}}(E)$] and multiply $I_0(E)$ by the relevant geometrical and time integration factors. The number of counts $C_0(E)$ accumulated by the detector over a given integration period $T_{\mathrm{obs}}$ is thus
\begin{align}
C_0(E) 	= I_0(E) \cdot G \cdot T_{\mathrm{obs}} \cdot \Delta E.
\end{align}
During primary observation, $G$ for the asteroid and the CXB are those given in Table \ref{tab:SpectGraspInputs}. During the calibration period for the CXB, however, $G$ is that for the whole spectrometer 
(i.e. $4.24~\mathrm{cm^2 Sr}$) and instead of $T_{\mathrm{obs}}$, we have $T_{\mathrm{CXB}}$.

\begin{figure}[htpb]
\begin{center}
\includegraphics[width=.65\textwidth]{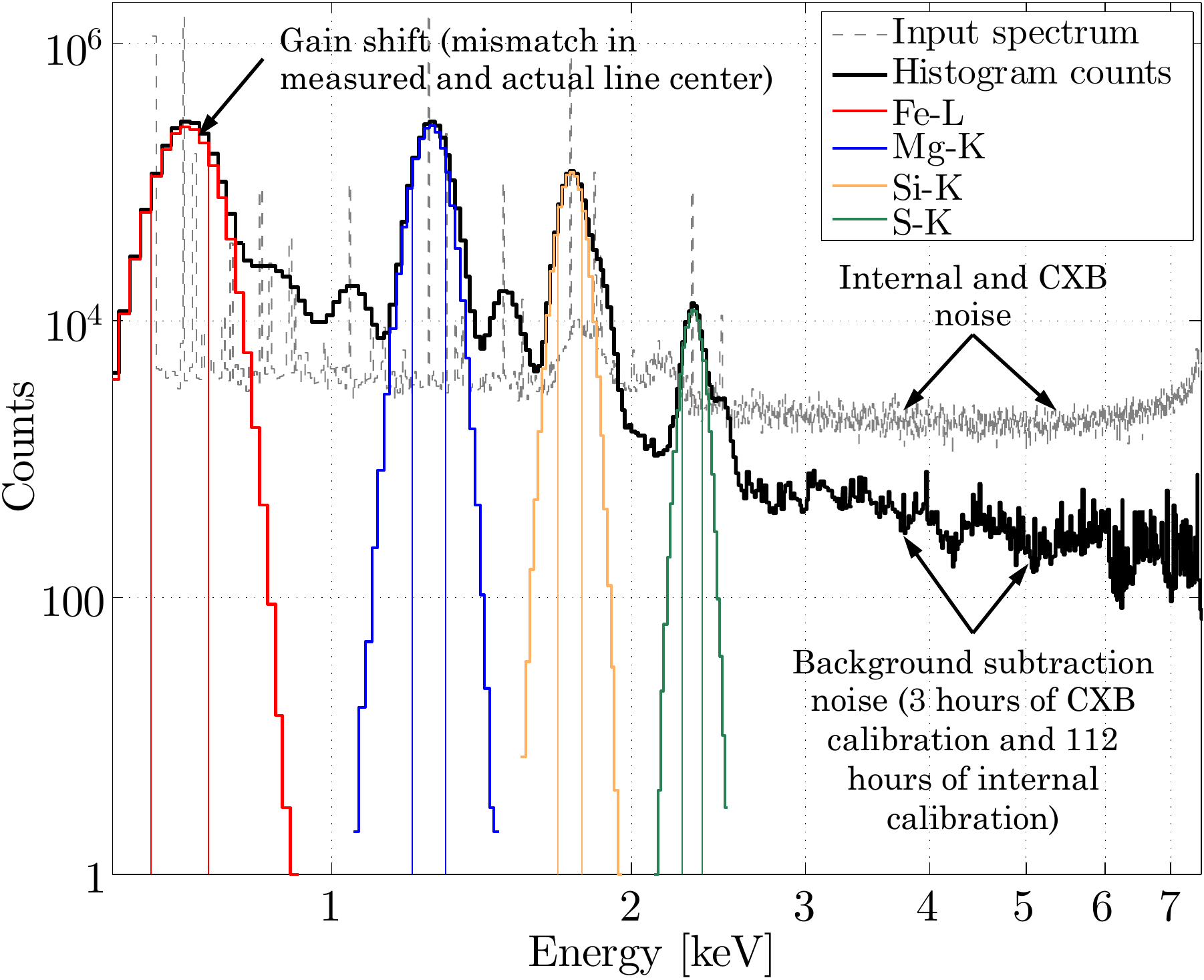}
\caption{Example histogram of detector at temperature of $-60~\mathrm{^{\circ}C}$, 
where these results follow from our preliminary model in Fig. \ref{fig:Spect_Comp}.}
\label{fig:Det60_hist}
\end{center}
\end{figure}

$C_0(E)$ will be reduced due to the quantum efficiency of the detector, and the resulting count distribution $C_1(E)$ is given by
\begin{align}
C_1(E) 	&= C_0(E)\cdot Q(E) \nonumber \\
	    	&= I_0(E) \cdot Q(E) \cdot G \cdot T_{\mathrm{obs}} \cdot \Delta E.
\end{align}
In Fig. \ref{fig:Spect_Comp}, we show how Bennu's modeled spectrum compares to the CXB and internal background, plotting for each $C_1(E)/T_{\mathrm{obs}}/\Delta E$. Consider a function  $\mathrm{Poisson}[C(E)]$ which
takes as an input a number of counts for a given energy $C(E)$ and outputs a Poisson-distributed random number from a distribution whose mean is $C(E)$. Applying Poisson statistics to $C_1(E)$ then gives
\begin{align}
C_2(E)	&= \mathrm{Poisson}\left[C_1(E)\right] \nonumber \\
      		&= \mathrm{Poisson}\left[I_0(E) \cdot Q(E) \cdot G \cdot T_{\mathrm{obs}} \cdot \Delta E\right]. 
\end{align}
The effect of the detector state upon the spectrum is accounted for by imposing an effective broadening upon each count value in the spectrum, the broadening having the shape of a Gaussian with a given $\mathrm{FWHM}$. For the CCDs, $\mathrm{FWHM} = \mathrm{FWHM}(E,T)$, where $T$ is the detector temperature and $E$ is photon energy. This broadening will not have the shape of a precise Gaussian, however, and to simulate the stochastic nature of the broadening, we generate a random distribution sampled from a Gaussian of given energy and $\mathrm{FWHM}$, with the total number of counts given by $C_2(E)$. Let $\mathrm{Gaussian}\left[E,C(E),\mathrm{FWHM}(E,T)\right]$ denote the generic Gaussian function that takes as an input the energy $E$, the counts at that energy $C(E)$, and the $\mathrm{FWHM}(E,T)$. Then the distribution of counts $C_3(E)$ is given by the convolution of $\mathrm{Gaussian}$ and $C_2(E)$:
\begin{align}
C_3(E) = C_2(E) \ast \mathrm{Gaussian}\left[E,C_2(E),\mathrm{FWHM}(E,T)\right].
\end{align}
A histogram is then generated by binning $C_3(E)$ into the required number of bins. Consider a generic binning function that takes as inputs a counts profile $C(E)$ over what may be regarded as a continuous energy range $E \in [0.5~\mathrm{keV},7.5~\mathrm{keV}]$ and bins it into a new profile $C'(E)$ over an energy range $E'$, where 
\begin{align*}
E' = \left\{0.5~\mathrm{keV}, 0.5~\mathrm{keV} + \Delta E, 0.5~\mathrm{keV} + 2\Delta E,...,7.5~\mathrm{keV}\right\}. 
\end{align*}
Denote this function $C'(E') = \mathrm{Binning}\left[C(E),E;E'\right]$. Then the final histogram profile $C_3'(E')$ over the binned energies is given by
\begin{align}
C_3'(E') = \mathrm{Binning}\left[C_3(E),E;E'\right]. 
\end{align}
In Fig. \ref{fig:Det60_hist} we show a simulated histogram for a detector temperature $T = -60~\mathrm{^\circ C}$. For reference, the spectral features associated with our lines of interest are shown in thick colored lines. Noise subtraction (see Sec. \ref{sec:BackSub} below) has been applied. In Fig. \ref{fig:Quiet_Fit}, we show the simulated histogram for the quiet Sun (solid magenta line), along with the idealized spectrum from which it is derived (dotted red line).

\subsection{Data Processing} 
In this section, we detail our methods for processing the simulated spectrometer and SXM data (right hand and bottom side of Fig. \ref{fig:Spect_Pipeline}). We first begin by detailing the process of noise background subtraction.  Next, we discuss how we perform histogram counts for all our lines of interest. We then discuss the method for reconstructing the Solar spectrum and, finally, how we generate calibration curves to map from histogram count ratios back to elemental abundance weight ratios. 

\subsubsection{Spectrometer Background Subtraction}\label{sec:BackSub}
As noted, asteroid spectral features (in particular, sulfur) are sensitive to noise from the CXB and from the instrument itself. As a result, REXIS devotes periods of time for both CXB and internal noise calibration. The data gathered during the calibration period are used to subtract out sources of noise from the data product. To simulate the noise subtraction procedure, we consider the total observed histogram counts $C_3'(E')$, which includes both the internal and CXB signal. We then simulate the calibration period data. Let internal counts as a function of energy be given by $C'_{\mathrm{int}}(E')$ and CXB counts by $C'_{\mathrm{CXB}}(E')$, where we have assumed that Poisson statistics and binning have been performed on each, and that the counts have been accumulated over the calibration periods, $T_{\mathrm{int}}$ and $T_{\mathrm{CXB}}$, respectively, for each. Then the procedure for background subtraction is to scale each calibration count value up so that the integration time matches that of the primary observation. Thus, the spectrum that we consider after accounting for noise subtraction is given by
\begin{align}
C_3'(E') - C'_{\mathrm{int}}(E')\frac{T_{\mathrm{obs}}}{T_{\mathrm{int}}}   
		- C'_{\mathrm{CXB}}(E') \frac{T_{\mathrm{obs}}}{T_{\mathrm{CXB}}}.
\end{align}
In Fig. \ref{fig:Det60_hist}, we see the effect of noise subtraction. The thin, dashed line shows the input spectrum with CXB and internal noise especially prominent at higher energies. After subtracting out CXB and internal background, there remains some high frequency noise (right side of Fig. \ref{fig:Det60_hist}) since Poisson statistics are included for both the simulated observation and calibration data. We note again that, since REXIS performs its CXB calibration with a sky observation in the absence of Bennu, $C'_{\mathrm{CXB}}(E')$ is calculated with a grasp given by that for the entire spectrometer.

\subsubsection{Spectrometer Line Counting} 
The quantity of various elements present in Bennu's regolith are determined by the strength of the corresponding spectral features and hence counts in the spectrometer histogram. When counting, for a given line center, we consider all the counts within the $\mathrm{FWHM}$ of a Gaussian centered about that line center. As discussed above, the $\mathrm{FWHM}$ is a function of both line center energy $E$ and detector temperature $T$. Onboard calibration data, which gives us $\mathrm{FWHM}$ at $5.9~\mathrm{keV}$, and pre-flight test data allow us in principle to estimate the detector $\mathrm{FWHM}$ for each CCD frame that is processed. For the purposes of our simulations here, we assume complete knowledge of $\mathrm{FWHM}$. Furthermore, as we demonstrate below (Sec. \ref{sec:Accs}), our results are relatively insensitive to $\mathrm{FWHM}$. We do not assume that we know the actual line centers with complete certainty. By employing gain drift (see Sec. \ref{sec:GainDrift} above), we allow for the misidentification of the line centers. In Fig. \ref{fig:Det60_hist}, we indicate the counting zones for the lines of interest by vertical lines. \textit{All} counts within a given $\mathrm{FWHM}$ counting zone are considered to come from that line of interest. 
Therefore, there will naturally be contamination within each zone from both the continuum background, noise sources, and other lines. During our simulations, we therefore keep track of the ideal, expected count number from each line in addition to those we count directly from the histogram. The error between the two affects how well we are able to reconstruct weight ratios from these counts. More details on the counting scheme and the definition of ``accuracy'' are given in Appendix \ref{sec:DefofAcc}.

\subsubsection{Solar Spectrum Reconstruction} 
\begin{figure}[htpb]
\begin{center}
\fbox{\includegraphics[width=0.75\textwidth]{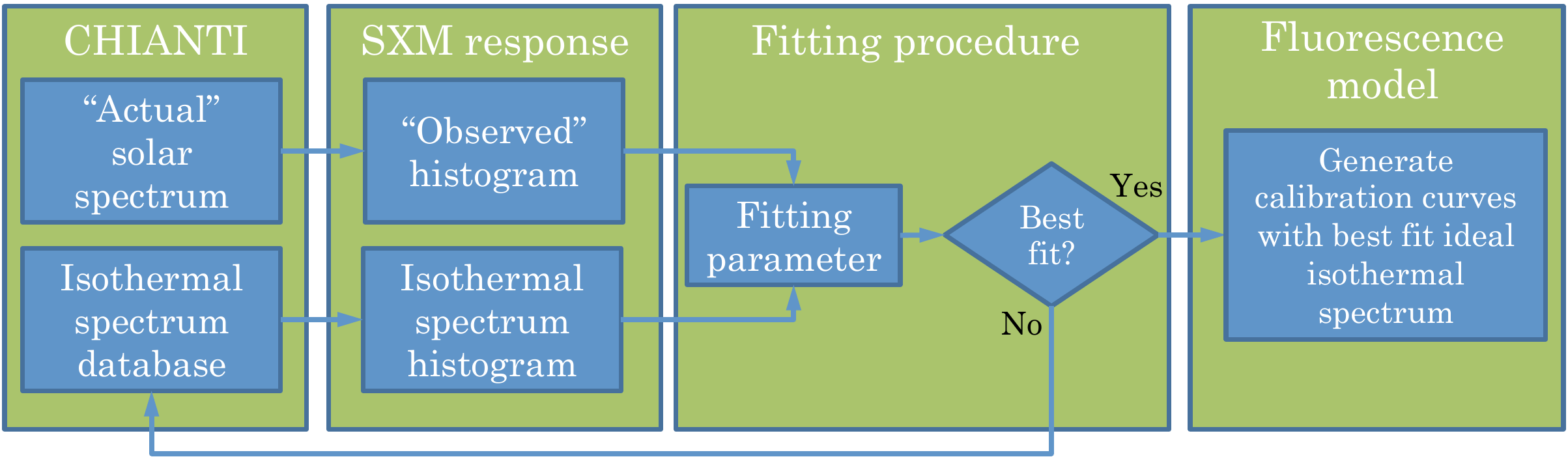}}
\caption[SXM modeling flow.]{Solar X-ray Monitor (SXM) modeling flow. Here, we start with the ``actual'' Solar spectrum, 
which is multithermal (or equivalently, differential emission measure convolved) and which we model using the CHIANTI
atomic database\cite{CHIANTI_1,CHIANTI_2} and the SolarSoftWare package \cite{SSW_pack}. This spectrum is 
convolved with the SXM instrument response function, which subsequently creates a synthetic Solar X-ray histogram data
product. We rely on the fact that, in the quiet state, the Solar X-ray spectrum can be approximated as isothermal. We hence 
generate a database of isothermal spectra, which we likewise convolve with the SXM response function and then fit to the 
response-convolved quiet Sun histogram. The isothermal spectrum corresponding to the best fit isothermal histogram 
is then used to generate calibration curves.}
\label{fig:SXM_flow}
\end{center}
\end{figure}

\begin{figure}[htpb]
\begin{center}
\includegraphics[width=0.8\textwidth]{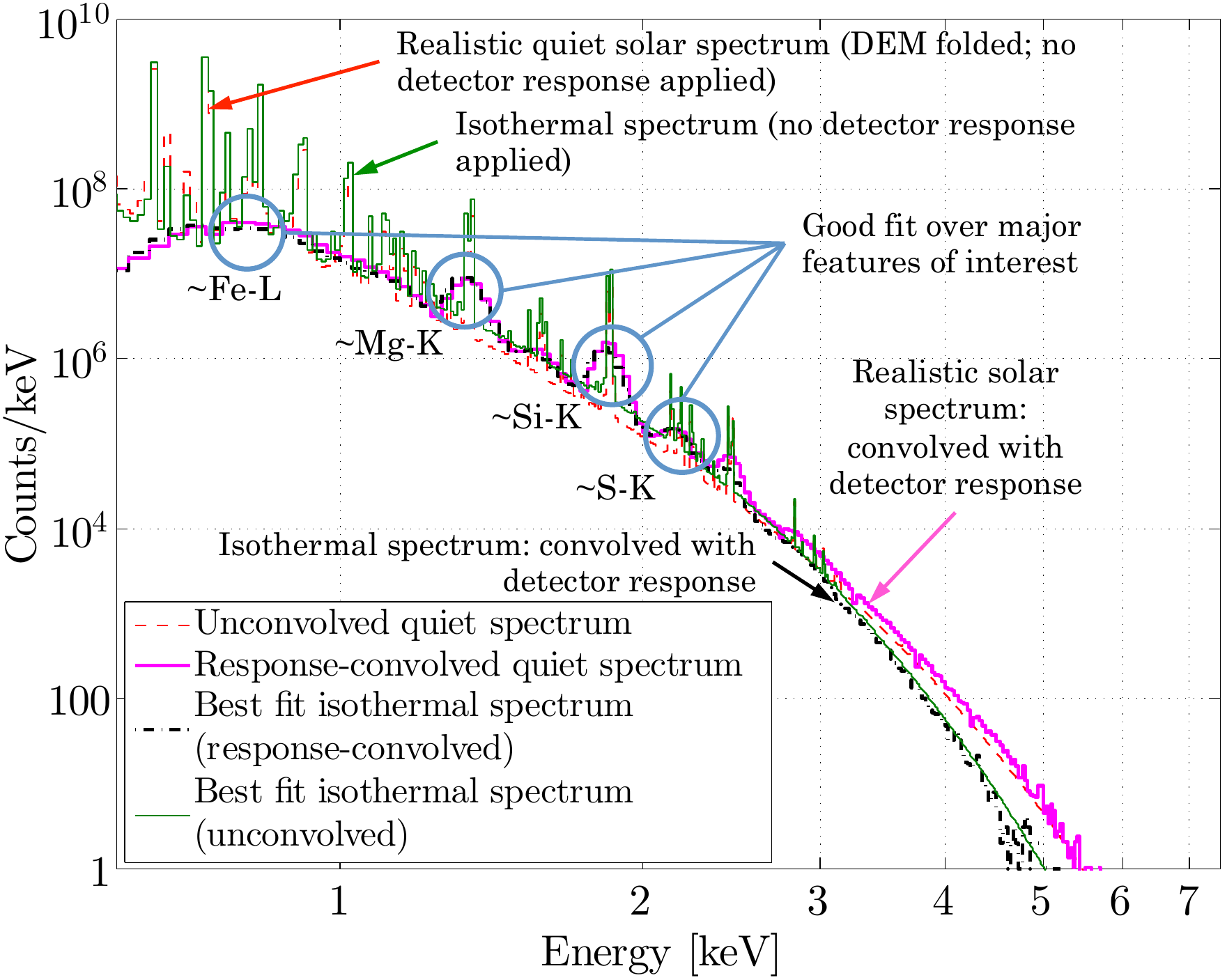}
\caption{Fitting of quiet Solar spectra in the isothermal approximation. Example of fitting procedure for a simulated quiet 
Sun. First, a DEM-folded, quiet Sun Solar spectrum model is convolved with the SXM detector response function, which 
simulates ground data for the Solar state. Then, we search for a best fit isothermal spectrum and convolve that with the 
detector response function. The isothermal spectrum whose detector response has the best fit with that of the realistic
spectrum is then used to generate the calibration curves (see Fig. \ref{fig:CalCurves}). The histogram shown is over 
32 s, while in this particular case, the isothermal best fit is a $3.1~\mathrm{MK}$ spectrum.
}
\label{fig:Quiet_Fit}
\end{center}
\end{figure}

The method by which we use the SXM histogram to reconstruct the Solar spectrum is shown schematically in Fig. \ref{fig:SXM_flow}. First we take the quiet Sun spectrum (see Sec. \ref{sec:SolarSpec}) and convolve it with the detector response to generate a synthetic ``observed'' histogram (two blue boxes on the upper left of Fig. \ref{fig:SXM_flow}). Then we utilize a database of isothermal spectra which we generate beforehand and convolve those with the known detector response to generate isothermal spectrum-derived histograms (boxes on the lower left). We use the observed histogram and those generated from the database to determine a best fit. The unconvolved isothermal spectrum whose convolved form provides the best fit is used in our later analysis. In Fig. \ref{fig:Quiet_Fit}, we summarize the results of the fitting procedure. In Fig. \ref{fig:Quiet_Fit}, we
see that there are good fits over the energy ranges corresponding to our elements of interest. 

We have found that using a linear combination of isothermal spectra when fitting against the observed Solar spectrum can provide better fits. This result is to be expected since the realistic quiet spectrum is indeed, via integration over the DEM, a linear combination of isothermal spectra. For simplicity here, however, we focus only on single-temperature fits. The quality of the fit depends also on the characteristics of the isothermal spectrum database. These spectra are dependent on factors such as the coronal elemental abundance model employed. While we do not claim to have explored the full model space of elemental abundance models available, we have ensured that the abundance models used for both the DEM-convolved realistic spectrum and that from which the isothermal spectrum database is derived are distinct and randomly chosen.

\subsubsection{Calibration Curve Generation: Mapping Count Ratios to Weight Ratios}
\begin{figure}[htpb]
        \centering
        \begin{subfigure}[t]{0.483\textwidth}
                \includegraphics[width=\textwidth]{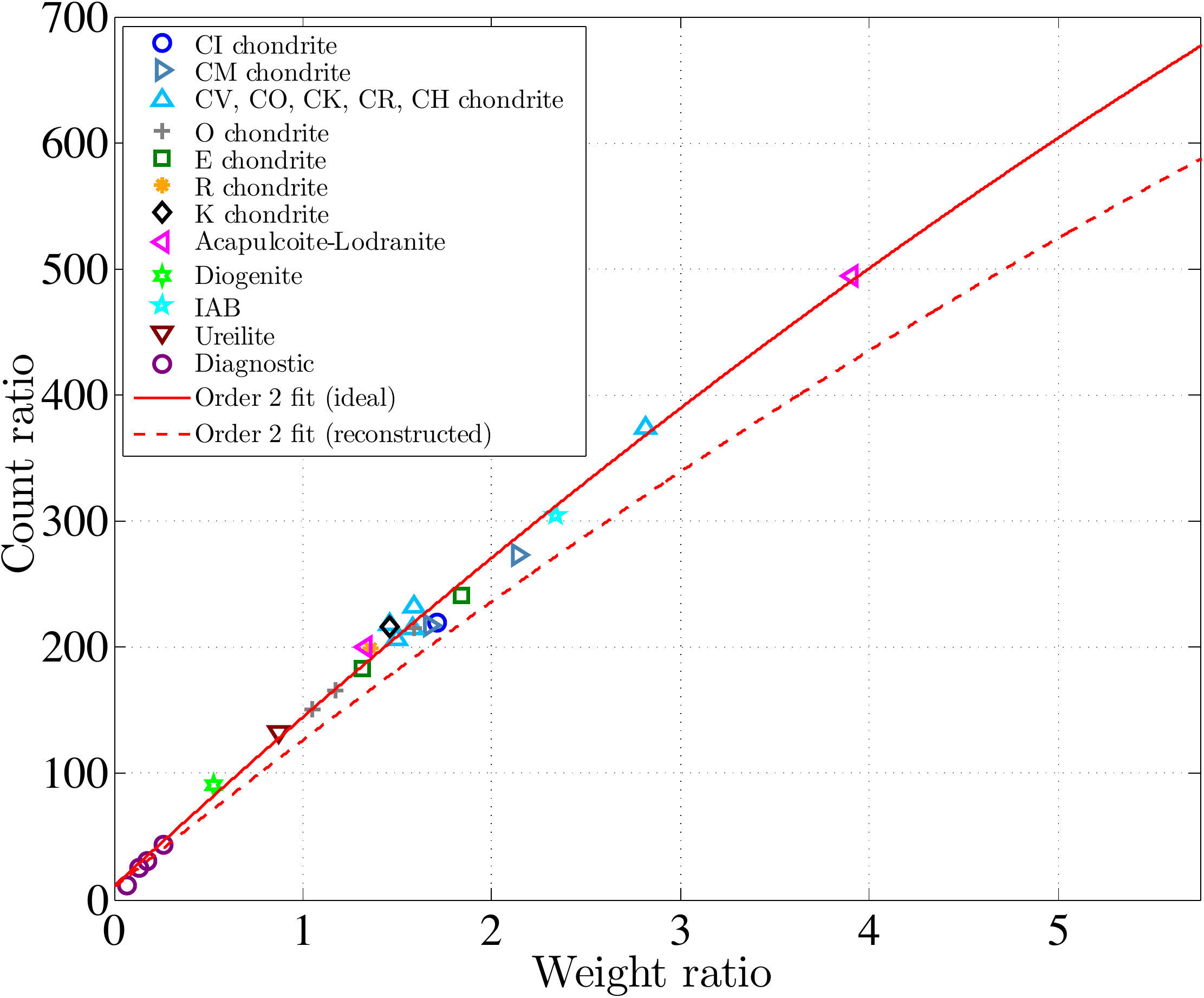}
                \caption{Fe/Si weight ratio to count ratio calibration curve.}
                \label{fig:FeSi_CC}
        \end{subfigure}%
        \quad 
        \begin{subfigure}[t]{0.471\textwidth}
                \includegraphics[width=\textwidth]{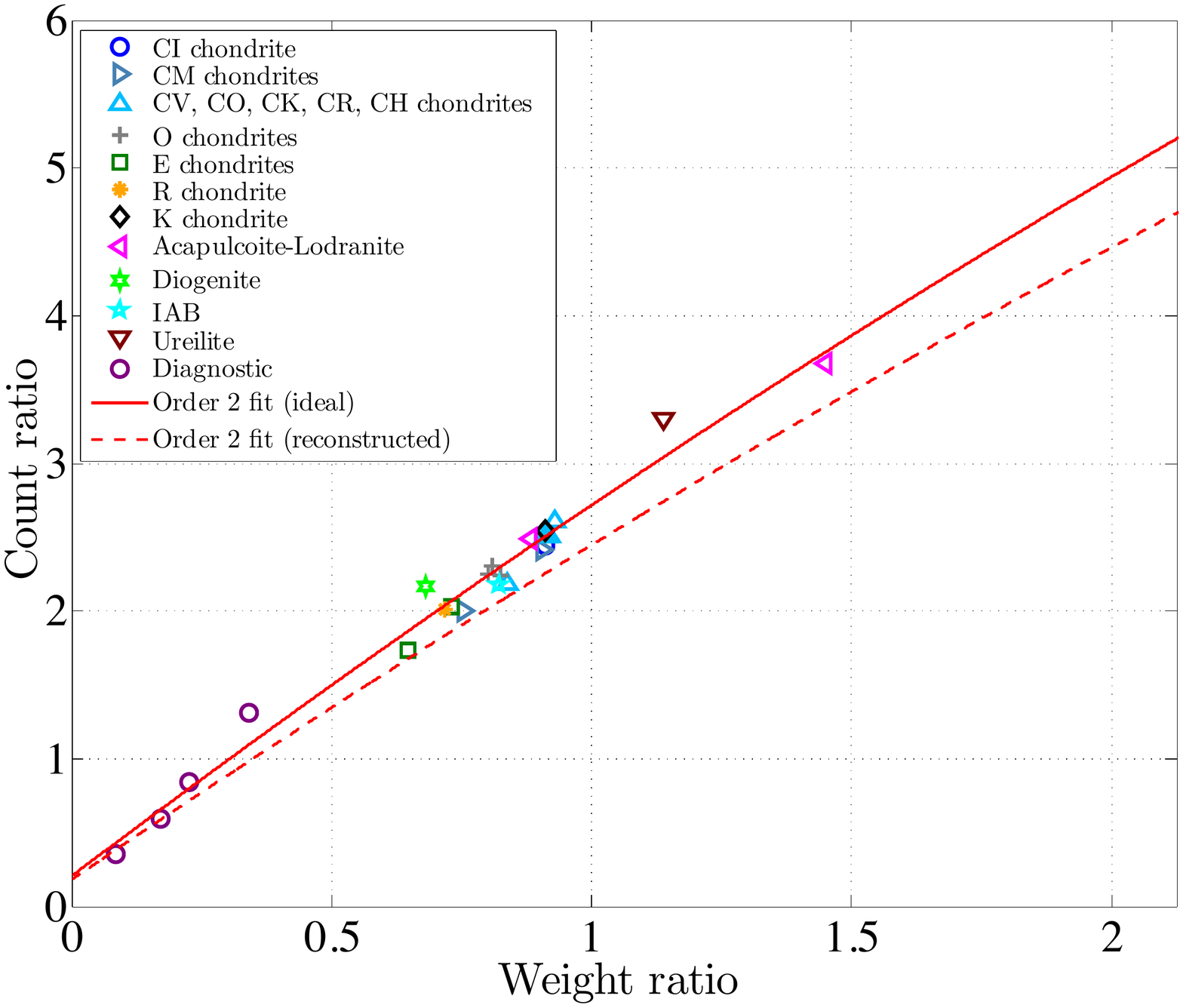}
                \caption{Mg/Si weight ratio to count ratio calibration curve.}
                \label{fig:MgSi_CC}
        \end{subfigure}
        \quad
       	\begin{subfigure}[t]{0.483\textwidth}
                \includegraphics[width=\textwidth]{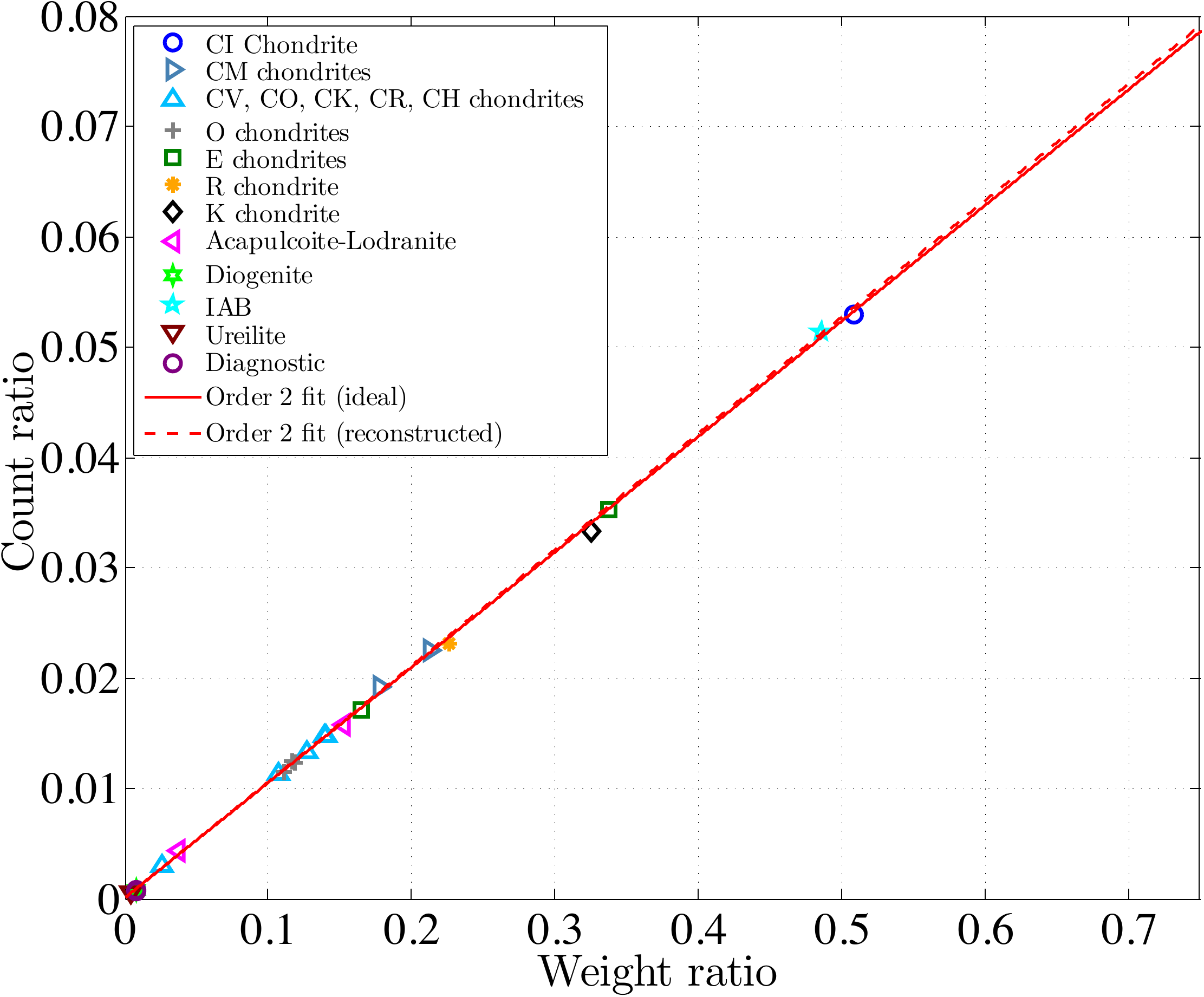}
                \caption{S/Si weight ratio to count ratio calibration curve.}
                \label{fig:SSi_CC}
        \end{subfigure}
        \caption{Weight ratio to count ratio calibration curves for the elemental abundance ratios of interest. Calibration 
        curves are a function of weight ratio and Solar spectrum. Curves indicate second order fits to simulated spectra of 
        asteroids with the same composition as the major meteorite types indicated (see Table \ref{tab:CalInputs}). Several
        artificial, ``diagnostic'' compositions have also been included to improve the fidelity of the fit. The solid 
        curve indicates spectra generated using the DEM-folded quiet Solar spectrum. The dotted curve indicates spectra
        generated using the isothermal best fit (see Fig. \ref{fig:Quiet_Fit}). The baseline weight ratios and expected count 
        ratios for a CI chondritic-like regolith are indicated by the blue circles. 
        }
        \label{fig:CalCurves}
\end{figure}

For a given Solar state, in order to make the transition from histogram count ratios to elemental abundance ratios, we make use of so-called calibration curves \cite{LimNitt1}. Calibration curves map, for a given Solar state, elemental abundance ratios to ideal count ratios. To generate calibration curves, we take the unconvolved Solar spectrum and simulate asteroid spectra corresponding to a wide range of meteorite compositions. The range of elemental abundances afforded by this range allows us to consider realistic weight ratios that may be expected from Bennu. The simulated spectra allow us to determine the expected count ratios for a given weight ratio, which then allows us to map histogram counts back to elemental abundances. In Table \ref{tab:CalInputs}, we detail the various meteoritic compositions used to generate the  calibrations curves for our elements of interest. In Fig. \ref{fig:CalCurves}, we show calibration curves for our elemental ratios of interest. Various meteorite groups are indicated by the symbols. Solid lines indicate second-order fits to the realistic quiet Solar spectrum, while the dotted line indicates the fit based on the reconstructed Solar spectrum, which we discuss further below. For simplicity, we show the symbols
only for the realistic quiet Sun fit and omit those for the reconstructed fit. The baseline weight ratios and expected count ratios for a CI chondrite-like regolith are indicated by the blue circles.

\begin{table}[htpb]
\caption[Summary of inputs for calibration curves.]{Summary of inputs for calibration curves. Values are taken from 
Lodders and Fegley\cite{PSC} unless marked with ``$^{*}$'', in which case the data comes from 
Nittler, \textit{et al.}\cite{NittlerData}. If ``---'' appears, weight percent values were not available in the reference for 
that element, and the remainder of the percent balance was allocated to that element for the sake of the ideal 
asteroid spectrum simulation. Since O is not one of our elements of interest, the development of the calibration 
curves is not incumbent upon accurate knowledge of O.}
	\centering
		\begin{tabular}{|c||c|c|c|c|c|c|}\hline
\multirow{2}{*}{Class} &\multicolumn{6}{c|}{Weight percent by element} \\  \cline{2-7}
	    		& O     	&Mg    	&Al     	&Si     	&S      	&Fe   	\\ \hline \hline
CI    			&46.4   	&9.70 	&0.865 	 &10.64 	&5.41 	&18.2  	\\ \hline 
CM    		&43.2   	&11.5 	&0.130 	 &12.70 	&2.70 	&21.3  	\\ \hline 
CM$^{*}$		&38.92  	&8.99 	&1.334  	&11.916 	&2.122 	&25.466	\\ \hline 
CV    		&37.0   	&14.3 	&1.680  	&15.70 	&2.20 	&23.5	\\ \hline 
CO    		&37.0   	&14.5 	&1.400  	&15.80 	&2.20 	&25.0	\\ \hline 
CK    		&---    	&14.7 	&1.470  	&15.80 	&1.70 	&23.0	\\ \hline 
CR    		&---    	&13.7 	&1.150  	&15.00 	&1.90 	&23.8	\\ \hline 
CH    		&---    	&11.3 	&1.050  	&13.50 	&0.35 	&38.0	\\ \hline 
H    			&35.70  	&14.1  	&1.06   	&17.1 	&2.0  	&27.2 	\\ \hline 
L    			&37.70  	&14.9  	&1.16   	&18.6 	&2.2  	&21.75	\\ \hline 
LL    		 	&40.00  	&15.3  	&1.18   	&18.9 	&2.1  	&19.8 	\\ \hline 
EH    		&28.00  	&10.73 	&0.82   	&16.6 	&5.6  	&30.5 	\\ \hline 
EL    		 	&31.00  	&13.75 	&1.00   	&18.8 	&3.1  	&24.8 	\\ \hline 
R    			&---    	&12.9  	&1.06   	&18.0 	&4.07 	&24.4 	\\ \hline 
K    			&---    	&15.4  	&1.30   	&16.9 	&5.5  	&24.7 	\\ \hline 
Acap.$^{*}$	&---    	&15.6  	&1.20   	&17.7 	&2.7  	&23.5 	\\ \hline 
Lod.$^{*}$ 	&25.858 	&16.299 	&0.0952 	&11.248 	&0.4257 	&43.92  	\\ \hline 
Dio.$^{*}$ 	&20.42  	&16.528 	&1.00   	&24.28  	&0.204 	&12.729	\\ \hline 
IAB$^{*}$  	&26.62  	&11.92 	&1.31   	&14.48 	&7.04 	&33.9   	\\ \hline 
		\end{tabular}
	\label{tab:CalInputs}
\end{table}

\newpage
\section{Results}\label{sec:Results}
In this section, we detail our results based on the simulations and analysis presented in the previous sections. We first show how, assuming perfect knowledge of the Solar state, the count ratios we measure map back to elemental abundance weight ratio errors with respect to CI chondrite-like baseline composition. Next, based on these count ratio errors, we present the required observation times to achieve statistical significance on our measurements. Finally, we present a qualitative discussion of how the error generated during Solar spectrum reconstruction develops a permissible error space in which we interpret our results.

\subsection{Weight Ratio Accuracy}\label{sec:Accs}
\begin{figure}[htpb]
        \centering
        \begin{subfigure}[b]{0.68\textwidth}
                \includegraphics[width=\textwidth]{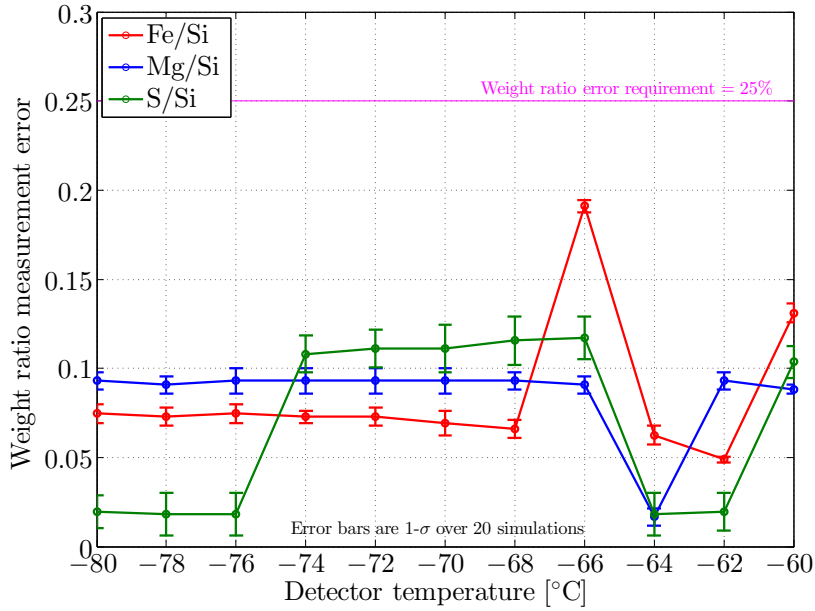}
                \caption{Weight ratio error as a function of detector temperature.}
                \label{fig:FWHM}
        \end{subfigure}%
	\\~
        \begin{subfigure}[b]{0.68\textwidth}
                \includegraphics[width=\textwidth]{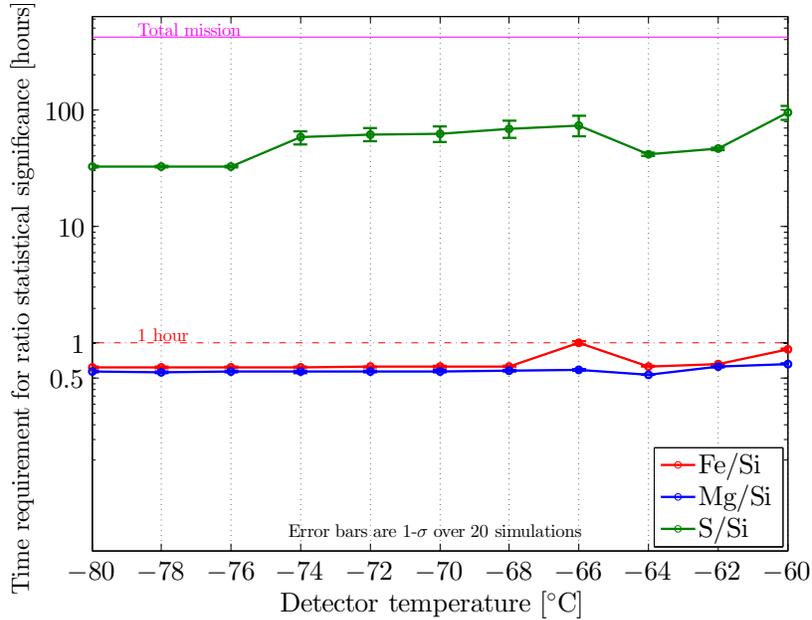}
                \caption{Observation time requirement to achieve statistical significance.}
                \label{fig:QE}
        \end{subfigure}
        \caption{Spectral simulation results assuming perfect knowledge of the Solar state. In the left panel, we show the 
        weight ratio error for our line ratios of interest as a function of detector temperature. The 25\% requirement is indicated
        by the magenta line. For all temperatures, we are able to meet our requirement with margin. In the right panel, we
        show the observation time required to achieve statistical significance on our spectral measurements. The requirement
        that we achieve significance within the allotted mission time is shown by the magenta line. The observation time is
        derived from the count rates from the lines of interest, CXB, and internal noise, and from the difference
        between predicted accuracy error and required accuracy error. S/Si, which is most susceptible to noise from the 
        internal background and CXB, requires the greatest observation time. Error bars are $1\sigma$ over 20 simulations. 
                }
        \label{fig:DetSumm}
\end{figure}

Using the calibration curves given in Fig. \ref{fig:CalCurves}, we can map the errors incurred by our histogram counting procedure into subsequent errors in weight ratio. In general, since the relationship between counts and regolith weight is roughly linear, the correspondence between count ratio error and weight ratio error is also roughly linear. In Table \ref{tab:Rattempsum}, we list the weight ratio errors for our required detector temperature ($T = -60~\mathrm{^{\circ} C}$) and our current best prediction for the detector temperature ($T \sim -80~\mathrm{^{\circ} C}$). In all cases, the predicted error is less than the requirement. The weight ratio errors over the range of temperatures between $T = -60~\mathrm{^{\circ} C}$ and  $-80~\mathrm{^{\circ} C}$ are shown on the left panel of Fig. \ref{fig:DetSumm}. The error bars in the figure represent the error spread over 20 simulations at each detector temperature, with each simulation incorporating the effect of factors such as Poisson statistics, gain drift, and noise subtraction. We note that, over most of the temperatures, there is not necessarily a degradation of performance with increasing detector temperature, as we might naively expect due to the decrease in spectral resolution. The relative insensitivity of our spectral performance on detector temperature (or equivalently $\mathrm{FWHM}$) is primarily due to the fact that taking count ratios effectively cancels some of the effect of this systematic error present in each of the individual lines. In Fig. \ref{fig:Nittler}, we indicate with magenta ellipses the accuracy error due to these systematic effects at $T = -60~\mathrm{^{\circ} C}$. 

\begin{table}[htpb]
	\centering
	\caption[Summary of REX-3 systematic \textit{count} ratio error.]{Summary of REX-3 systematic accuracy error
	performance. We show the accuracy error for each elemental abundance ratio of interest at two different detector
	temperatures. The detector temperature of $-60~\mathrm{^{\circ}C}$ represents the required detector operating
	temperature, while $-80~\mathrm{^{\circ}C}$ represents the current best prediction for the detector temperature
	at the time of writing. 
	} 
\begin{tabular}{cc|c||c|c|c|} \cline{3-6}
	&&\multirow{2}{*}{Ratio}	& Predicted accuracy	& Requirement 		& Margin  			\\
	&&					& error [$\%$] 			& [$\%$] 			& (Requirement/Prediction)	\\ \hhline{--====}
\multicolumn{1}{ |c| }{\multirow{6}{*}{\rotatebox[origin=c]{90}{Det. Temp. $T$}}} & \multicolumn{1}{ |c|| }{\parbox[t]{2mm}{\multirow{3}{*}{\rotatebox[origin=c]{90}{\small{$-60~^{\circ}$C}}}}} 	
																								&Fe/Si 	& 13.7	& $\leq 25\%$ 	& 1.8   	\\ \cline{3-6}
\multicolumn{1}{ |c| }{}&\multicolumn{1}{ |c|| }{}																&Mg/Si	& 9.1    	& $\leq 25\%$  	& 2.7	  	\\ \cline{3-6}
\multicolumn{1}{ |c| }{}&\multicolumn{1}{ |c|| }{}																&S /Si 	& 10.3     	& $\leq 25\%$ 	& 2.4  	\\ \cline{2-6}
\multicolumn{1}{ |c| }{}&\multicolumn{1}{ |c|| }{\parbox[t]{2mm}{\multirow{3}{*}{\rotatebox[origin=c]{90}{\small{$-80~^{\circ}$C}}}}}&Fe/Si 	& 7.9		& $\leq 25\%$ 	& 3.2 	\\ \cline{3-6}
\multicolumn{1}{ |c| }{}&\multicolumn{1}{ |c|| }{}																&Mg/Si	& 9.7    	& $\leq 25\%$  	& 2.6	  	\\ \cline{3-6}
\multicolumn{1}{ |c| }{}&\multicolumn{1}{ |c|| }{}																&S /Si 	& 3.0     	& $\leq 25\%$ 	& 8.3  	\\ \hline
\end{tabular}
	\label{tab:Rattempsum}
\end{table}

\subsection{Observation Time}\label{sec:ObsTime}
The results in the above section represent systematic error. That is, they represent errors intrinsic to the behavior of the instrumentation itself. We must also consider statistical error to account for the stochastic nature of photon emission and place a statistical significance on our expected results. In order to account for statistical error, we consider the quadratic difference between our expected \textit{count} ratio error and allowed count ratio error. Then, assuming Poisson statistics, based on the count rates within each energy range of interest from both the fluorescent lines and noise sources, we determine the required observation time to achieve an $N\sigma$ statistical significance level. Here, we choose $N = 3.5$, corresponding to $> 99\%$ confidence. A detailed calculation of the above, as well as expected count rates, is given in Appendix \ref{sec:StatTime} and Table \ref{tab:Countsum}, respectively. Our required observation times for the two detector temperatures discussed above are given in Table \ref{tab:ObsTimeSum}, while those for the range of temperatures in between are given in the right panel of Fig. \ref{fig:DetSumm}. Again, in all cases, we are able to achieve our required performance with margin. As noted above, the S/Si ratio, which is most subject to the effect of CXB and internal noise, requires the greatest amount of observation time to achieve statistical significance. The magenta error ellipses shown in Fig. \ref{fig:Nittler}
thus have a $3.5\sigma$ statistical confidence associated them.

\begin{table}[htpb]
	\centering
	\caption[Summary of REX-6, observation time requirement]{Summary of REX-6 observation time requirement
	expected performance for detector temperature of $-60~\mathrm{^{\circ}C}$ and $-80~\mathrm{^{\circ}C}$. 
	The observation time is based on obtaining sufficient photon statistics to achieve $3.5\sigma$ confidence in the 
	accuracy results.} 
\begin{tabular}{cc|c||c|c|c|} \cline{3-6}
	&&\multirow{2}{*}{Ratio}	& Observation time for $3.5\sigma$	& Requirement 	& Margin  					\\ 
	&&					&  confidence  [hours] 			& [hours]		& (Requirement/Prediction)	\\ \hhline{--====}
\multicolumn{1}{ |c| }{\multirow{6}{*}{\rotatebox[origin=c]{90}{Det. Temp. $T$}}} & \multicolumn{1}{ |c|| }{\parbox[t]{2mm}{\multirow{3}{*}{\rotatebox[origin=c]{90}{\small{$-60~^{\circ}$C}}}}} 	
																								&Fe/Si 	& 0.9    	& $\leq 420$ 	& 467   	\\ \cline{3-6}
\multicolumn{1}{ |c| }{}&\multicolumn{1}{ |c|| }{}																&Mg/Si	& 0.7     	& $\leq 420$  	& 600 	\\ \cline{3-6}
\multicolumn{1}{ |c| }{}&\multicolumn{1}{ |c|| }{}																&S /Si 	& 108     	& $\leq 420$ 	& 3.9		\\ \cline{2-6}
\multicolumn{1}{ |c| }{}&\multicolumn{1}{ |c|| }{\parbox[t]{2mm}{\multirow{3}{*}{\rotatebox[origin=c]{90}{\small{$-80~^{\circ}$C}}}}} &Fe/Si 	& 0.6    	& $\leq 420$ 	& 700	\\ \cline{3-6}
\multicolumn{1}{ |c| }{}&\multicolumn{1}{ |c|| }{}																&Mg/Si	& 0.6     	& $\leq 420$  	& 700  	\\ \cline{3-6}
\multicolumn{1}{ |c| }{}&\multicolumn{1}{ |c|| }{}																&S /Si 	& 33     	& $\leq 420$ 	& 12.7		\\ \hline
\end{tabular}
	\label{tab:ObsTimeSum}
\end{table}

\subsection{Calibration Curves and Mapping Errors}\label{sec:ErrorSpace}
\begin{figure}[ht]
\begin{center}
\includegraphics[width=.8\textwidth]{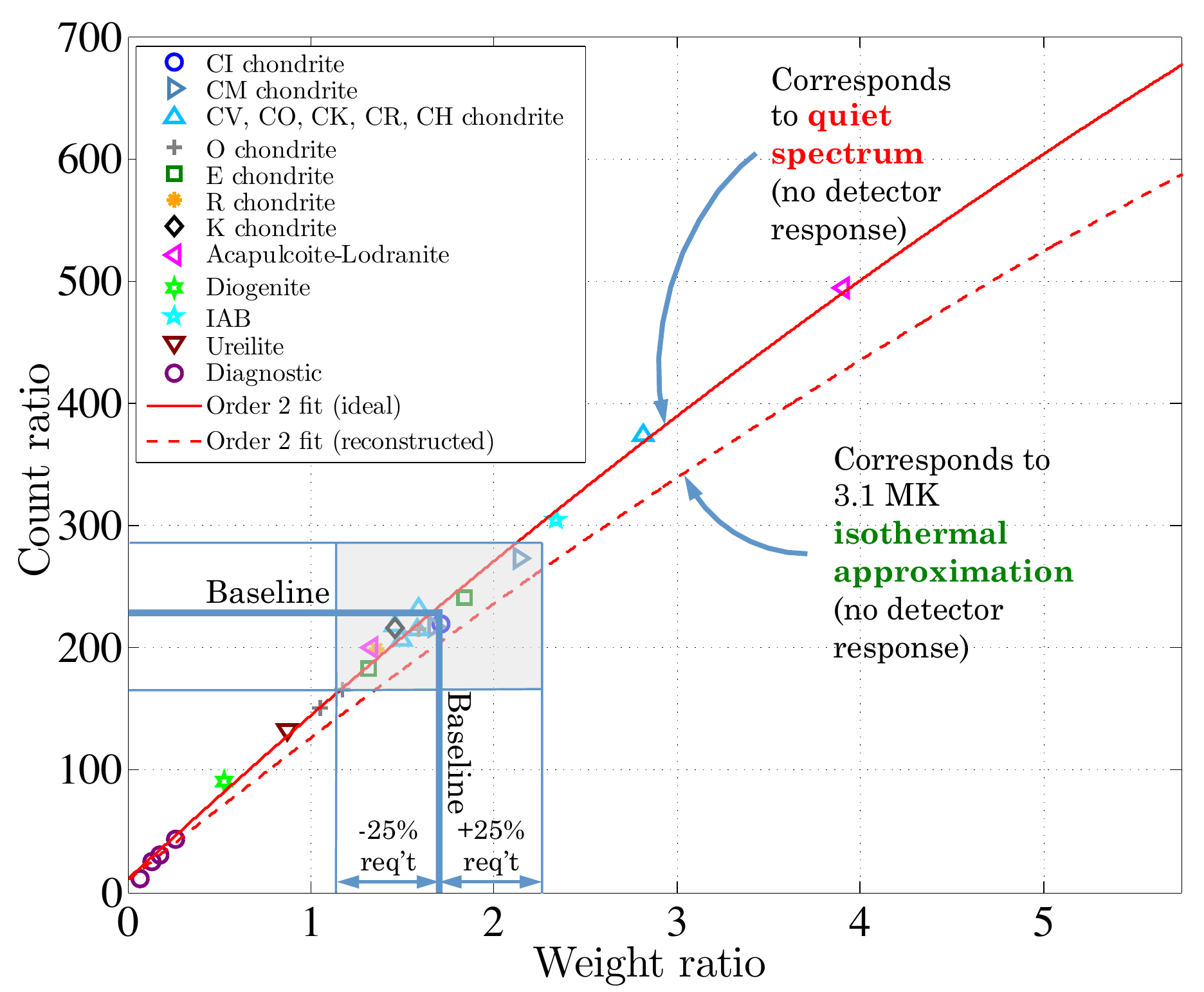}
\caption{Calibration curve error space. The solid red line is the calibration curve generated by the quiet Sun (red line in 
Fig. \ref{fig:Quiet_Fit}). This is the calibration curve that would be generated if we had perfect knowledge of the Solar state. 
The dashed red line is the calibration curve generated by the best single temperature fit based on simulated SXM data
(green line in Fig. \ref{fig:Quiet_Fit}).}
\label{fig:Cal_Curve_Error_Space}
\end{center}
\end{figure}

In the results above, we have assumed perfect knowledge of the Solar state in mapping count ratio errors to weight ratio errors. Hence, we have used the solid red calibration curves shown in Fig. \ref{fig:CalCurves}. In reality, we will have to use the reconstructed Solar spectrum-derived calibration curves (dotted red lines in Fig. \ref{fig:CalCurves}) in order to perform the mapping. From the calibration curves shown in Fig. \ref{fig:CalCurves}, it is clear that the difference between the curves based on the actual and reconstructed Solar spectra are within the error of the fit itself over the weight ratio ranges of interest, so that we 
cannot claim a truly meaningful difference between the quality of each fit. While we have not accounted for this effect in the results presented above, we demonstrate graphically in Fig. \ref{fig:Cal_Curve_Error_Space} the error space that develops from reconstructing the Solar spectrum. Fig. \ref{fig:Cal_Curve_Error_Space} shows how, in the most extreme case (Fe/Si) the calibration curves diverge for the different input Solar spectra. The lines marked ``baseline'' map to a CI chondrite-type composition under the actual Solar spectrum. The 25\% identification requirement then places a range of permissible errors about this baseline (the shaded grey area). So long as the combined misidentification of the Solar spectrum and the count ratio error does not exceed the bounds set by the shaded region, REXIS can still achieve its objectives. In our case, for the $T=-60~\mathrm{^\circ C}$ case, the Fe/Si error is the most marginal ($13.7\%$ error), although it still falls within the required performance region. Since Fe/Si is not subject to the the same statistical fluctuation as e.g. S/Si, statistical significance on Fe/Si can still easily be achieved within the allotted observation time.

\section{Discussion and Conclusions}
In the previous sections, we have presented the methodology and results of spectral performance modeling of REXIS. We have shown, by simulating Solar and asteroid X-ray spectra, the subsequent data product, and the data processing, how well REXIS can be expected to identify Bennu as a CI chondrite analog. We have shown that our two primary requirements---that REXIS is capable of identifying a baseline CI chondrite meteorite analog for Bennu to within 25\% and that it can accomplish this within the allotted mission observation time---are attainable with margin. 
\subsection{Future Work}
This work represents the first step in understanding REXIS's science performance in Spectral Mode, and there are numerous opportunities to extend and refine this work. We summarize some of these below.
\subsubsection{Other Baseline Regolith Compositions for Bennu} 
Throughout this work, we have assumed a baseline CI chondrite-like regolith composition for Bennu. In reality, ground measurements have suggested a possible CM chondrite-like composition. While we should not expect any substantial difference in expected performance if we assumed a CM-type composition, it is worthwhile to consider the possibility that the baseline composition of the regolith is something radically different (for instance, achondritic). 
\subsubsection{Higher Order Observational Effects}
We have assumed here that the orbit is perfectly circular and that Bennu is a perfect sphere, e.g. we do not take into account surface roughness. However, it is possible to use the Bennu shape model\cite{ShapeModel} and OSIRIS-REx orbit for REXIS science operations in order to model the effect of both shape and orbit to a higher level of fidelity. 

\subsubsection{Active Sun Modeling and Reconstruction}
We have assumed throughout this work that the Sun is in a quiet state, which it will be for the majority of REXIS's science operations. However, the Sun will occasionally flare, creating a higher flux and harder (i.e. greater intensity at higher energies) spectrum. This in turn will substantially affect Bennu's spectrum. Performing a similar analysis to the above, but assuming a flare Sun, should be carried out. The flare Sun, however, cannot be approximated as isothermal, although a two-temperature model may suffice [see Appendix \ref{sec:SolarModelSec} and also Lim and Nittler (2009)\cite{LimNitt1}]. 

\subsubsection{Improved SXM Modeling}
In this work, we have assumed a relatively simple geometry and instrument response function for the SXM. Much as we have done for the spectrometer, it is possible to compute higher fidelity values for the SXM grasp
and, with continued testing, better characterize the SXM response function in general. 

\subsubsection{Radiation Damage}
Preliminary work has suggested that under OSIRIS-REx's expected radiation environment, degradation in CCD spectral resolution due to non-ionizing radiation damage still permits REXIS to meet its science objectives. This is in part due to the presence of the radiation cover \cite{RadDam,Harrison,RadDamChandra} and the fact that REXIS is primarily concerned with the measurement of count ratios, which reduces the effect of spectral degradation on weight ratio reconstruction error. (Sec. \ref{sec:Accs}). However, continued characterization of the REXIS CCDs should allow for a more definite characterization of REXIS's radiation environment and the effect of radiation damage on spectral performance (especially as a function of time), which we have not accounted for here. 

\subsubsection{Internal Background}
The internal background spectrum we have used here is scaled from Chandra data. In the future, a more accurate model making use of the actual REXIS geometry should be employed to determine the fluorescent signature of the REXIS structure in response to X-rays from Bennu and the CXB. In this case, a simulation framework such as GEANT4\cite{GEANT} can be used to determine the intensity of X-ray emission from REXIS itself incident upon the CCDs.

\subsubsection{Further Exploration of Error Space}
While our discussion of the error space in Sec. \ref{sec:ErrorSpace} was somewhat qualitative, it is worthwhile to be more quantitative about our approach. Furthermore, we may continue to characterize how each of the model inputs discussed in Sec. \ref{sec:Method} (such as molecular contamination, the OBF, and the spectral resolution) independently affect spectral performance.

\section*{Acknowledgments}       
The authors would like to thank Dr. Steve Kissel of the MIT Kavli Institute for CCD test data, Beverly LaMarr of the MIT Kavli Institute for discussions regarding radiation damage to CCDs, and Dr. Lucy Lim of NASA GSFC and Dr. Ben Clark of the Space Science Institute for many helpful discussions and suggestions concerning asteroid spectroscopy. This work was conducted under the support of the OSIRIS-REx program through research funds from Goddard Space Flight Center.

\newpage
\appendix    
\section{Modeling Bennu}\label{sec:AsteroidSec}
\subsection{Fluorescence}
We model Bennu as a sphere of $280~\mathrm{m}$ radius, with OSIRIS-REx viewing Bennu in a terminator orbit at $1~\mathrm{km}$ from the asteroid barycenter. Denote a point on Bennu by $P$, the center of REXIS's detector plane by $R$, and the center of mass of the Sun by $S$. Let the angle between surface normal at $P$ and $\overline{SP}$ be given by $\psi_{\mathrm{in}}$, and that between the surface normal at $P$ and $\overline{PR}$ by $\psi_{\mathrm{out}}$. Then the intensity for the $k^{\mathrm{th}}$ fluorescence line as measured by REXIS is given by \cite{JenkinsQXS}
\begin{align}
I_k(E_k) = \frac{\Omega_{\astrosun}}{\Omega_B }\int_{\mathrm{Bennu}}
\frac{d\Omega_B Q_{k}(E_k)}{4\pi \Delta E \sin \psi_{\mathrm{in}}}
\int_{E_k}^{\infty}\frac{I_{\odot}(E)dE}{\sum_{j}W_j\left[\mu_j(E)\csc\psi_{\mathrm{in}} 
+ \mu_j(E_k)\csc\psi_{\mathrm{out}}\right]}
\end{align}
where $Q_{k}(E_k)$ is a factor that encompasses the probability and quantum yield associated with emission of the $k^{\mathrm{th}}$ line.  $\Omega_{\astrosun}$ is the solid angle subtended by the Sun with respect to Bennu, $\Delta E$ is an arbitrarily-chosen energy bin, and $\Omega_B$ is the solid angle subtended by Bennu with respect to REXIS. If we assume all incident Solar X-rays are parallel to one another (valid since the Sun can effectively be considered a point source with respect to Bennu), then $\psi_{\mathrm{in}}$, $\psi_{\mathrm{out}}$, and $d\Omega_B$ can be related to one another by straightforward geometry, and an integration over the entire surface area of Bennu amounts to an integration over all the $\psi_{\mathrm{out}}$ within the REXIS field of view.

If $W_k$ is the weight fraction of the element associated with the $k^{\mathrm{th}}$ line, $r_k$ is the jump ratio, $\omega_k$ is the fluorescence yield, and $f_k$ is the fraction of the series to which $k$ belongs that is devoted to $k$, we have 
\begin{align}
Q_{k}(E_k) = W_k  \frac{r_k - 1}{r_k}\omega_k f_k.
\end{align}
Since fluorescent intensity is monochromatic, the total contribution to the spectrum from fluorescence is the sum of all the $I_k$. An example of the calculation of the line/probability factor for the Fe-K series is given in Table \ref{tab:FeKSeries}. 

\begin{table}[htp]
	\centering
	\caption[Series information for Fe-K.]{Series information for Fe-K and calculation of associated line probability/yield 
	factor, $Q_{k}(E_k)$.}
\begin{tabular}{|c|c|c|c|c|c|c|} \hline
Edge/Series 			& $r_k$                 		& $\omega_k$ 			& Line 		& $E_k$ [eV]	& $f_k$			 		& $Q_{k}/W_k$ 			\\ \hline \hline
\multirow{6}{*}{Fe-K} 	& \multirow{6}{*}{$0.351$}	& \multirow{6}{*}{$7.893$} & K$\alpha_3$	& $6,267.40$  	& $2.76096\times 10^{-4}$  	& $7.649\times 10^{-4}$		\\ \cline{4-7}
					&					&    					& K$\alpha_2$	& $6,392.10$	& $2.94023\times 10^{-1}$	& $8.145\times 10^{-1}$  		\\ \cline{4-7}
					&					&  					& K$\alpha_1$	& $6,405.20$	& $5.80277\times 10^{-1}$	& $1.608$	              			\\ \cline{4-7}
					&					&     					& K$\beta_3$	& $7,059.30$ 	& $4.25566\times 10^{-2}$	& $1.179\times 10^{-1}$		\\ \cline{4-7}
					&					&     					& K$\beta_1$	& $7,059.30$ 	& $8.21556\times 10^{-2}$	& $2.276\times 10^{-1}$		\\ \cline{4-7}
					&					&     					& K$\beta_5$	& $7,110.00$	& $7.12115\times 10^{-4}$	& $1.973\times 10^{-3}$		\\ \hline
\end{tabular}
	\label{tab:FeKSeries}
\end{table}
\subsection{Coherent scattering}
The continuous spectrum due to coherent scattering is given by
\begin{align}
I_{\mathrm{scattering}}(E) = \frac{1}{\Omega_B }\int_{\Omega} \frac{d\sigma}{d\Omega}I_{\odot}(E)N_A d\Omega_B,
\end{align}
where $N_A$ is Avogadro's number and the integration effected over the solid angle $\Omega$. The differential scattering cross section $d\sigma/d\Omega$ is given by
\begin{align}
\frac{d\sigma}{d\Omega} = \frac{r_e^2}{4\pi}\left(1 - \cos^2\theta \right)\left|F\right(E,\theta)|^2,
\end{align}
where $r_e$ is the classical electron radius, given by $2.82\times 10^{-15}~\mathrm{m}$, and where $|F(E,\theta)|^2$ the modulus squared of the (complex) atomic form factor $F(E,\theta)$ for the element in question, dependent upon both the energy of the incident radiation and the scattering angle $\theta \equiv \left|\psi_{\mathrm{in}} + \psi_{\mathrm{out}}\right|$. Incoherent scattering, being at least an order of magnitude
smaller than coherent scattering for all energies and elements of interest, is not considered. The total intensity of radiation emitted by the asteroid (in units of $\mathrm{photons/Sr/s/eV/cm^2}$) is given by the sum of fluorescent and scattered radiation: 
\begin{align}
I(E) = \sum_{k}I_k(E_k) + I_{\mathrm{scattering}}(E).
\end{align}

\section{Modeling the Solar Spectrum}\label{sec:SolarModelSec}
Here we briefly discuss how the Solar corona is modeled in order to determine its X-ray spectrum. Denote the power per unit volume emitted by a plasma undergoing an atomic transition from quantum states $j \rightarrow i$ by $P_{ij}$, and the wavelength (or equivalently, the energy) associated with this transition by $\lambda_{ij}$. Then the intensity $I_{\odot}(\lambda_{ij})$ of the radiation at the surface of the body of interest (say Bennu) for this transition is given by
\begin{align}
I_{\odot}(\lambda_{ij}) = \frac{1}{4\pi R^2}\int_{V}P_{ij}dV,
\end{align}
where $R$ is the distance from the Sun to Bennu and $V$ is the plasma volume. $P_{ij}$ can be written \cite{SolarCorona}
\begin{align}
P_{ij} = 0.8 A_{X} G(T,\lambda_{ij})\frac{hc}{\lambda_{ij}}N_e^2,
\end{align}
where $N_e$ is the local electron density, $A_X$ is the abundance of the $X$th element with respect to hydrogen, and $G$ is the so-called contribution function (not to be confused with the grasp), which is a function of both the plasma temperature $T$ (not to be confused with the detector temperature) and the wavelength $\lambda_{ij}$ associated with the transition.

$P_{ij}$ is dependent upon both temperature and electron density, and it is possible to decompose it into density and temperature dependent parts. We define the ``differential emission measure'', or $\mathrm{DEM}$, as follows
\begin{align}\label{eq:DEMdef}
\int_V N_e^2 dV = \int_T \mathrm{DEM}(T) dT,
\end{align}
so that 
\begin{align}
P_{ij} = 0.8 A_{X} G(T,\lambda_{ij})\frac{hc}{\lambda_{ij}} \mathrm{DEM}(T)\frac{dT}{dV}.
\end{align}
On the left hand side of Eq. \eqref{eq:DEMdef}, the integration is effected over the plasma volume; on the right hand side, over the possible coronal temperatures. Defining
\begin{align}
\phi(T,\lambda) \equiv \sum_{X}\sum_{ij} 0.8 A_{X} G(T,\lambda_{ij})\frac{hc}{\lambda_{ij}}
\end{align}
(the first sum extending over all species and the second sum extending over all transition pairs), we have, using Eq. \eqref{eq:DEMdef},
\begin{align}\label{eq:DEMtoI}
I_{\odot}(\lambda) = \int_T \phi(T,\lambda) \mathrm{DEM}(T) dT.
\end{align}
Instead of integrating over all possible coronal temperatures, in certain cases, it is possible to consider only those coronal temperatures at which the emission measure is greatest. For quiet Solar regions, it has been found that the emission measure peaks at one value, while the $\mathrm{DEM}$ for active regions tends to peak at two values. Thus we may write
\begin{align}\label{eq:TempModels}
I_{\odot}(\lambda) &= \phi(T_1,\lambda)\mathrm{EM}(T_1) \hspace{3.325cm}\textrm{(quiet Sun)} \\
I_{\odot}(\lambda) &= \phi(T_1,\lambda)\mathrm{EM}(T_1) + \phi(T_2,\lambda)\mathrm{EM}(T_2) \hspace{0.5cm}\textrm{(active Sun)}
\end{align}
where $\mathrm{EM}(T)$ is a single temperature emission measure, which for all practical purposes here amounts to a simple numerical prefactor.

\section{Definition of Accuracy}\label{sec:DefofAcc}
\begin{figure}[H]
\begin{center}
\includegraphics[width=0.7\textwidth]{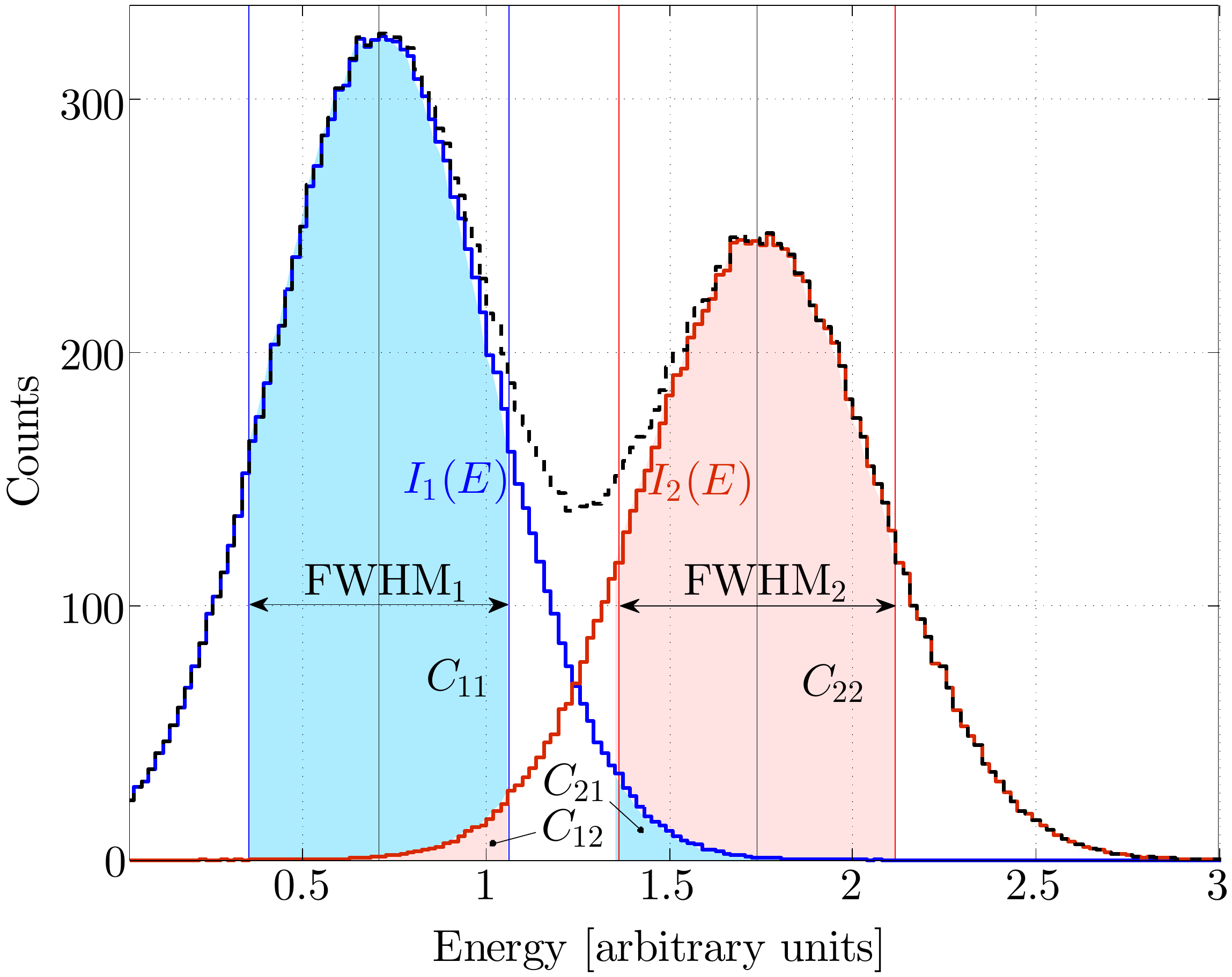}
\caption[Example of line contamination in the simplified case of only two lines.]{Example of line contamination in the simplified case of only two lines, $I_1(E)$ (blue) and $I_2(E)$ (red); their sum is in black. The counts $C_{11} + C_{12}$ account for all the counts under the sum in the $\mathrm{FWHM}$ zone of line 1 and the counts $C_{22} + C_{21}$ are all the counts in the $\mathrm{FWHM}$ zone of line 2. $C_{11}$ are the counts that are due to line 1 in the line 1 $\mathrm{FWHM}$ zone, and $C_{22}$ are the counts due to line 2 in the line 2 $\mathrm{FWHM}$ zone, while $C_{12}$ is the contamination from line 2 into the FWHM zone of 
line 1, and $C_{21}$ the contamination of line 1 into the $\mathrm{FWHM}$ zone of line 2. 
}
\label{fig:ContamExam}
\end{center}
\end{figure}		

REXIS's spectral performance requirement states that the reconstructed weight ratios of the asteroid regolith be within a certain percent of that of the baseline composition. REXIS itself can only measure count ratios, and we use the calibration curves to make the correspondence between count ratios and weight ratios. For convenience, we denote this function $\mathrm{CC}: \textrm{count ratio} \rightarrow \textrm{weight ratio}$, with the inverse mapping $\mathrm{CC}^{-1}:  \textrm{weight ratio}\rightarrow \textrm{count ratio}$. 

Fig. \ref{fig:ContamExam} demonstrates REXIS's counting procedure in forming count ratios. For simplicity, the figure shows only two spectral features $I_{1}$ and $I_{2}$, shown in blue and red, respectively. First, we define counting zones centered about each line center. The width of these zones are given by the full-width half-maximum $\mathrm{FWHM}$ of the Gaussian centered at each energy. All the counts within each zone are considered to be from the respective line, although in reality, there will be some contamination from other spectral features. Thus in our simplified case, the total number of counts in $\mathrm{FWHM}_1$ will include contributions from $I_{1}$, denoted $C_{11}$, and those from $I_2$, denoted $C_{12}$. Likewise, the total number of counts in $\mathrm{FWHM}_2$ will include contributions from $I_2$ (given by $C_{22}$) and $I_1$ (given by $C_{21}$). The total count ratio of the first feature to the second $\rho_{1/2}$ is then given by 
\begin{align}
\rho_{1/2} = \frac{C_{11} + C_{12}}{C_{22} + C_{21}}.
\end{align}
In the more general case applicable to REXIS, we have the total number of binned detector counts given by $C_3'(E')$, so that for the $\rho_{k/\mathrm{Si}}$ line ratio,
\begin{align}
\rho_{k/\mathrm{Si}} = \frac{\sum_{E_k - \mathrm{\tiny{FHWM}}_k/2}^{E_k + \mathrm{\tiny{FHWM}}_k/2}C'_3(E')}{\sum_{E_{\mathrm{Si}} - \mathrm{\tiny{FHWM}}_{\mathrm{Si}}/2}^{E_{\mathrm{Si}} + \mathrm{\tiny{FHWM}}_{\mathrm{Si}} /2}C'_3(E')}.
\end{align}
We map $\rho_{k/\mathrm{Si}}$ to the equivalent weight ratio $\varpi_{k/\mathrm{Si}}$ by using the calibration curve: $\varpi_{k/\mathrm{Si}} = \mathrm{CC}\left(\rho_{k/\mathrm{Si}}\right)$. The accuracy error is then calculated by comparing the measured weight ratio $\varpi_{k/\mathrm{Si}}$ with the regolith input weight ratio $\varpi_{k/\mathrm{Si},0}$ corresponding to a CI chondrite-like composition. If we denote the weight ratio error requirement by $\eta$, then we may write the requirement as
\begin{align}
\left|1 - \frac{\varpi_{k/\mathrm{Si}}}{\varpi_{k/\mathrm{Si},0}} \right| \leq \eta,
\end{align}
where $\eta = 0.25$. 

In some cases (e.g., Appendix \ref{sec:StatTime}), we may wish to consider only the errors in count ratios. In this case, we consider the expected count ratio from the lines of interest, ignoring contamination and 
other effects. If we denote this ratio by $\rho_{k/\mathrm{Si},0}$, we have
\begin{align}
\rho_{k/\mathrm{Si},0} = \frac{C_1\left(E_{k}\right)}{C_1\left(E_{\mathrm{Si}}\right)},
\end{align}
where care has been taken to ensure that the effect of quantum efficiency has been accounted for. Indeed, when the calibration curves are generated, $\rho_{k/\mathrm{Si},0}$ is calculated for a whole range of baseline compositions, and a second order fit performed on the various $(\varpi_{k/\mathrm{Si},0},\rho_{k/\mathrm{Si},0})$ pairs (Fig. \ref{fig:CalCurves}). The REXIS performance requirement in terms of count ratio is denoted by $\eta_C$, and is then given by
\begin{align}
\left|1 - \frac{\rho_{k/\mathrm{Si}}}{\rho_{k/\mathrm{Si},0}} \right| \leq \eta_C.
\end{align}
$\eta_C$ can be related to $\eta$ in a relatively straightforward way that is most clearly demonstrated graphically by means of the shaded region shown in Fig. \ref{fig:Cal_Curve_Error_Space}; it is given mathematically by
\begin{align}
\left|1 - \frac{\mathrm{CC}^{-1}\left[\left(1 - \eta\right) \times \varpi_{k/\mathrm{Si},0}\right]}{\mathrm{CC}^{-1}\left( \varpi_{k/\mathrm{Si},0}\right)}\right| = \eta_C.
\end{align}

\section{Statistical Error}\label{sec:StatTime}
Suppose for a given detector temperature that the total error in counts due to systematic error is $\Delta$, that the requirement is given by $\eta_{C}$, and that statistical error is given by $\sigma$. We suppose that the errors can be summed quadratically:
\begin{align}
\eta_{C} = \sqrt{\Delta^2 + \sigma^2},
\end{align}
so that
\begin{align}
\sigma = \sqrt{\eta_{C}^2 - \Delta^2}.
\end{align}

For the $k^{\mathrm{th}}$ line and an $N$ confidence level,
\begin{align}\label{eq:CountExp}
\frac{\sigma^2}{N^2} &= \frac{ \dot{N}_{k}/T_{\mathrm{tot}} + \dot{N}_{\mathrm{CXB},k}/T_{\mathrm{CXB}} + \dot{N}_{\mathrm{int},k}/T_{\mathrm{int}}}{\left(\dot{N}_{k} - \dot{N}_{\mathrm{CXB},k} - \dot{N}_{\mathrm{int},k}\right)^2} + \frac{ \dot{N}_{\mathrm{Si}}/T_{\mathrm{tot}} + \dot{N}_{\mathrm{CXB},\mathrm{Si}}/T_{\mathrm{CXB}} + \dot{N}_{\mathrm{int},\mathrm{Si}}/T_{\mathrm{int}}}{\left(\dot{N}_{\mathrm{Si}} - \dot{N}_{\mathrm{CXB},\mathrm{Si}} - \dot{N}_{\mathrm{int},\mathrm{Si}}\right)^2}.
\end{align}
$T_{\mathrm{CXB}}$ and $T_{\mathrm{int}}$ are the CXB and internal calibration times (see Sec. \ref{sec:BackSub}). $\dot{N}_{k}$, $\dot{N}_{\mathrm{CXB},k}$, and $\dot{N}_{\mathrm{int},k}$ refer respectively to the total count rates, CXB count rates, and internal background count rates within the $k^{\mathrm{th}}$ $\mathrm{FWHM}$ counting zone. More precisely, $\dot{N}_{k}$, $\dot{N}_{\mathrm{CXB},k}$, and $\dot{N}_{\mathrm{int},k}$ are given by $C_3'/T_{\mathrm{obs}}$, $C_{\mathrm{CXB}}'/T_{\mathrm{obs}}$, and $C_{\mathrm{int}}'/T_{\mathrm{obs}}$ summed over each $\mathrm{FWHM}$ zone. Rearranging Eq. \ref{eq:CountExp}, we get
\begin{align}
\frac{\sigma^2}{N^2} &=  \underbrace{\left[\frac{ \dot{N}_{k}}{\left(\dot{N}_{k} - \dot{N}_{\mathrm{CXB},k} - \dot{N}_{\mathrm{int},k}\right)^2} + \frac{ \dot{N}_{\mathrm{Si}}}{\left(\dot{N}_{\mathrm{Si}} - \dot{N}_{\mathrm{CXB},\mathrm{Si}} - \dot{N}_{\mathrm{int},\mathrm{Si}}\right)^2}\right]}_{\equiv L}\frac{1}{T_{\mathrm{tot}}} + \nonumber \\
& \underbrace{\frac{ \dot{N}_{\mathrm{CXB},k}/T_{\mathrm{CXB}} + \dot{N}_{\mathrm{int},k}/T_{\mathrm{int}} }{\left(\dot{N}_{k} - \dot{N}_{\mathrm{CXB},k} - \dot{N}_{\mathrm{int},k}\right)^2} + \frac{\dot{N}_{\mathrm{CXB},\mathrm{Si}}/T_{\mathrm{CXB}} + \dot{N}_{\mathrm{int},\mathrm{Si}}/T_{\mathrm{int}}}{\left(\dot{N}_{\mathrm{Si}} - \dot{N}_{\mathrm{CXB},\mathrm{Si}} - \dot{N}_{\mathrm{int},\mathrm{Si}}\right)^2}}_{\equiv R}.
\end{align}
With $L$ and $R$ defined as above, the total observation time $T_{\mathrm{obs}}$ required for $N\sigma$ confidence is given by
\begin{align}
T_{\mathrm{obs}} = \frac{L}{R- \sigma^2/N^2}.
\end{align}
A summary of the expected count rates within each $\mathrm{FWHM}$ zone are given in Table \ref{tab:Countsum}.
\begin{table}[htp]
	\centering
	\caption{Summary of 
	expected count rate with detector temperature $T = -60~\mathrm{^\circ C}$. $E_-$ and $E_+$ are the lower and 
	upper limits to the $\mathrm{FWHM}$ zone for each line, respectively. 
	$\dot{N}_{k}$ is the total number of counts in each $\mathrm{FWHM}$ zone (i.e., from Bennu and background),
	while $\dot{N}_{\mathrm{CXB}, k}$ and $\dot{N}_{\mathrm{int}, k}$ are CXB and internal count rates, respectively,
	in each zone.} 
\begin{tabular}{|c||c|c||c|c|c|c|c|} \hline
Line	& $E_-$ [keV]	&  $E_+$[keV] 	&$\dot{N}_{k}$	&$\dot{N}_{\mathrm{int},k}$   	&   $\dot{N}_{\mathrm{CXB},k}$\\ \hline \hline 
Fe-L	& 0.658		& 0.751		& 4.73		& 0.038 					&  3.73					\\ \hline 
Mg-K& 1.1830		& 1.3230 		& 2.64		& 0.0291					&  1.69					\\ \hline 
Si-K 	& 1.6620 		& 1.8139 		& 1.46		& 0.0745					&  0.948					\\ \hline 
S-K 	& 2.2250 		& 2.3910 		&0.967		& 0.0296					&  0.888					\\ \hline 
\end{tabular}
	\label{tab:Countsum}
\end{table}
\section{Calculating the Energy Resolution of the Detector}\label{sec:FWHMCalc}
To determine the energy and detector temperature dependence of the detector resolution $\mathrm{FWHM}$, we require two pieces of experimental data: $\mathrm{FWHM}$ as a function of energy $E$ at a fixed temperature 
$T_0$, and $\mathrm{FWHM}$ as a function of temperature $T$ at a fixed energy $E_0$. The two pieces of information can be combined then to determine the general dependence of $\mathrm{FWHM}$ on $E$ and $T$\footnote{Personal communication, M. Bautz.}:
\begin{align}\label{eq:FWHM_rearr}
\mathrm{FWHM}(E,T) = \sqrt{\mathrm{FWHM}^2(E,T_0) + \mathrm{FWHM}^2(E_0,T) - \mathrm{FWHM}^2(E_0,T_0)}
\end{align}
In the case of REXIS, energy resolution for the CCID-41 has been experimentally determined as a function of energy at $T_0 = -90~\mathrm{^{\circ}C}$\footnote{Personal communication, M. Bautz.}, and as a function of temperature at $E_0 = 5.89~\mathrm{keV}$\footnote{Personal communication, S. Kissel.}. These two pieces of information together allow us to use Eq. \ref{eq:FWHM_rearr} to generate Fig. \ref{fig:InstResInputs}.

\newpage
\bibliography{report}   
\bibliographystyle{spiebib}   

\end{document}